\documentclass[journal, compsoc]{IEEEtran}

\usepackage{latexsym}
\usepackage{multirow}
\usepackage{graphicx}
\usepackage{amsfonts,amssymb,amsmath}
\usepackage{hyperref}
\usepackage{algorithm}
\usepackage{algpseudocode}
\usepackage{xcolor}

\usepackage{tikz}
\usetikzlibrary{positioning}
\usetikzlibrary{bayesnet}

\ifCLASSOPTIONcompsoc
 \usepackage[nocompress]{cite}
\else
 \usepackage{cite}
\fi

\ifCLASSINFOpdf
\else
\fi

\begin{document}
\title{Bayesian causal inference in automotive software engineering and online evaluation}

\author{Yuchu Liu, David Issa Mattos, Jan Bosch, Helena Holmstr\"om Olsson, and Jonn Lantz 

\thanks{Y. Liu is with Volvo Cars,  405 31 G\"oteborg, Sweden and Chalmers University of Technology, 412 96 G\"oteborg, Sweden (email: \href{mailto:yuchu.liu@volvocars.com}{yuchu.liu@volvocars.com}).}

\thanks{D.I. Mattos is with Volvo Cars, 405 31 G\"oteborg, Sweden.}
\thanks{J. Bosch is with Chalmers University of Technology, 412 96 G\"oteborg.}
\thanks{H. H. Olsson is with Malm\"o University, 211 19 Malm\"o, Sweden.}
\thanks{J. Lantz is with Volvo Cars, 405 31 G\"oteborg, Sweden.}
}

\maketitle

\begin{abstract}
Randomised field experiments, such as A/B testing, have long been the gold standard for evaluating software changes.
In the automotive domain, running randomised field experiments is not always desired, possible, or even ethical.
In the face of such limitations, we develop a framework BOAT (\textbf{B}ayesian causal modelling for \textbf{O}bvserv\textbf{A}tional \textbf{T}esting), utilising observational studies in combination with Bayesian causal inference, in order to understand real-world impacts from complex automotive software updates and help software development organisations arrive at causal conclusions.
In this study, we present three causal inference models in the Bayesian framework and their corresponding cases to address three commonly experienced challenges of software evaluation in the automotive domain.
We develop the BOAT framework with our industry collaborator, and demonstrate the potential of causal inference by conducting empirical studies on a large fleet of vehicles. Moreover, we relate the causal assumption theories to their implications in practise, aiming to provide a comprehensive guide on how to apply the causal models in automotive software engineering.
We apply Bayesian propensity score matching for producing balanced control and treatment groups when we do not have access to the entire user base, Bayesian regression discontinuity design for identifying covariate dependent treatment assignments and the local treatment effect, and Bayesian difference-in-differences for causal inference of treatment effect overtime and implicitly control unobserved confounding factors.
Each one of the demonstrative case has its grounds in practise, and is a scenario experienced when randomisation is not feasible.
With the BOAT framework, we enable online software evaluation in the automotive domain without the need of a fully randomised experiment.



%
\end{abstract}

\begin{IEEEkeywords}
Automotive Software, Causal Inference, Software Engineering, Bayesian Statistics, Online Experimentation.
\end{IEEEkeywords}

\IEEEpeerreviewmaketitle

\section{Introduction \label{intro}}

Randomised online experiments, such as A/B testing, is a technique for evaluating the impact of software changes towards real users.
With the demonstrated success of online experiment implementation in Software-as-a-Service (SaaS) companies \cite{Deng2013, google2010, Xie2016}, the automotive domain starts to raise interest in adopting such a method, and even starts to conduct experiments for evaluating embedded software online \cite{Mattos2020, Liu2021, Liu2021a, Liu2021b,Giaimo2019, Mattos2022} 

Despite the known advantages and benefits, the automotive domain struggles to scale experimentation activities \cite{Liu2021a,Mattos2020, Giaimo2020}.
Some of the identified challenges are the limited number of users, limited capability of Over-The-Air (OTA) software deployment, strict user agreements, safety-critical and validation constraints among others. These limitations often create roadblocks that limit the scope and feasibility of conducting experiments in customer vehicles. To overcome these limitations, practitioners are looking to leverage collected observational data to understand the causal impact of a software change \cite{Xu2016}.

In our previous study \cite{Liu2021b}, we empirically applied and evaluated the use of causal modelling for software engineering, in which a Bayesian propensity score matching model is applied for generating balanced control and treatment groups in observational software testing.
However, as we have experienced further needs in causal inference in the automotive domain due to a number of limitations,
to address this need, this paper evaluates the use of three different Bayesian causal models for treatment effect inference from observational studies, applied to automotive software development. 
This work extends the method BOAT (\textbf{B}ayesian causal modelling for \textbf{O}bvserv\textbf{A}tional \textbf{T}esting) \cite{Liu2021b} to include the Bayesian propensity score matching model for producing balanced control and treatment groups, the Bayesian regression discontinuity design for identifying covariate dependent treatment assignment, and Bayesian difference-in-differences model for causal inference on treatment effect over time. 
While these models have been widely used in the frequentist setting in other domains of science (such as medicine \cite{Normington2019}, traffic and transport \cite{Li2020}, social studies \cite{Lee2008, Chib2015}), to the best of our knowledge, this is the first paper to apply and evaluate these models in the Bayesian setting and in the context of automotive software engineering.

We demonstrate the BOAT method with three cases from our industrial collaborations, utilising automotive embedded software deployed on real vehicles and users. Comparing with the existing literature, the contribution of this paper is three-fold.

\begin{itemize}
    \item We present an overview and discussion of causality in the potential outcomes framework applied in automotive software development, along with three illustrative studies that reinforce the need for Bayesian causal inference in software online evaluation.
    \item We demonstrate three different Bayesian causal inference models to assess the causal effects in their corresponding examples. These models are Bayesian regression discontinuity design, Bayesian difference-in-differences, and Bayesian propensity score matching.
    \item We relate the causal assumptions made in causal inference in relation to online observational studies conducted on automotive software, and we discuss their specific implications.
\end{itemize}


The rest of this paper is arranged as following.
We elaborate the importance of causality in automotive software engineering and present the theory and assumption in the potential outcome framework in \ref{background}. In Section \ref{theory}, we present the BOAT framework and our research method. The three Bayesian causal models and their related cases are described in Section \ref{BPSM}, \ref{BDID} and \ref{BRDD}. The discussion and conclusion are in Section \ref{diss} and \ref{conclude} respectively.
Moreover, we include an online appendix to share our Bayesian models and their inference.
\section{Background \label{background}}

In this section, we introduce the concept of randomised experimentation applied in software engineering, the potential outcomes framework, and its relevant theories. Moreover, we give an overview of Bayesian statistics and its inference.

\subsection{Randomised experimentation}
%
Randomised experiment methods, such as A/B testing, are common practises adopted by SaaS companies \cite{Deng2013, google2010, Xie2016}. 
In a two-level experiment, the sample group is split into control and treatment at random and exposed to different versions of the same software.
When an experiment is fully randomised, the outcome is independent of the treatment assignment, this is defined as exchangeability. In other words, the control and the treatment groups are interchangeable and do not have any preexisting differences, thus the observed outcome can only be caused by the treatment. Therefore, randomised experiments help us at establishing a causal relationship between the intervention and the outcome.

Causal knowledge helps us cope with change \cite{Pearl2009}.
Through data analytics, i.e., passive observation, we could compute a joint distribution of vehicle usage and performance -- however, such a distribution cannot inform us if a change in our product would or would not improve the product performance and user experience.
With intervention, randomised experimentation enables direct feedback from the users and helps organisations answer the "what-if" question to software changes.
Randomised experiment has long been the gold standard for evaluating software in an online and continuous manner. 

In recent years, there is an increasing interest of adopting online experimentation in the automotive domain but the ability to conduct large scale and fully randomised experiments is significantly more limited, as reported in \cite{Mattos2020, Liu2021a, Giaimo2020, Giaimo2019, Mattos2018}. 
The automotive domain faces many unique restrictions compared to SaaS companies, such as the number of hardware and software variants \cite{Mattos2020}, architecture restrictions \cite{Liu2021}, safety regulation constraints, number of vehicles available for experimentation \cite{Liu2021a}, driver consent, and the ability to frequently update software \cite{Mattos2020, Giaimo2020}. 
A combination of these challenges leads to many situations where a randomised experiment is not possible, such as in limited samples; desired, such as in highly regulated systems; or ethical, without explicit consent of the vehicle owners and users on the complete scope of the new software. 

When a randomised experiment is not feasible, there present confounding factors that can often cause a spurious correlation and hinder us from drawing causal conclusions \cite{Rosenbaum1983, Holland1986, Pearl2009}.
To address this issue, observational studies in combination with causal inference models and causal assumptions need to be applied. 
We will further discuss the empirical scenarios from the automotive sector and the underlying implications of causal assumptions on software observational testing in the following sections.

\subsection{The potential outcomes framework}

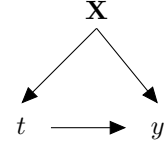
\begin{figure}[t]
\centering
\begin{tikzpicture}[
roundnode/.style={circle, draw=black!60, fill=white, very thick, minimum size=10mm},
roundedrect/.style={rectangle, rounded corners, minimum width=3cm, minimum height=1.5cm,text centered, draw=black}
]
\node      (maintopic)          {\begin{tabular}{c} $t$ \end{tabular}};
\node      (uppercircle)       [above=of maintopic, xshift=1.0cm] {$\mathbf{X}$};
\node      (rightcircle)       [right=of maintopic] {\begin{tabular}{c} $y$
\end{tabular} };

\draw[->] (uppercircle.south) -- (rightcircle.north);
\draw[->] (uppercircle.south) -- (maintopic.north);
\draw[->] (maintopic.east) -- (rightcircle.west);

\end{tikzpicture}
\caption{A simplified directed acyclic graph showing the relationships of treatment ($t$), target variable ($y$), and covariates ($\mathbf{X}$).}
\label{figure:DAG}
\vspace{-0.5cm}
\end{figure}

The potential outcomes framework \cite{Holland1986} describes causal inference from an intervention introduced in randomised experiments. In this section, we discuss the potential outcome from both experiments and observational studies, the later is extensively explored in studies such as \cite{Rosenbaum1983, Rubin1978}.
Potential outcome models quantify the treatment effect from the intervention introduced, or from known systematic differences between the control and treatment groups.

\textbf{A word on notation:} In this paper, we will use the following set of notations consistently throughout the different sections. In this paper, a lower case letter (such as $x$) denotes a vector, and when necessary, a superscript $x^T$ denotes a row vector of a random variable. The subscript $x_j$ denotes the length of the vector. The notation of $x = \{x_1, x_2, ... x_j\}$ indicates all the instances in the vector, and if a square bracket is used, we indicate the values in $x$ are bounded, e.g., $x = [0, 42]$. We use a bold upper case letter (such as $\mathbf{X}$, containing a set of vectors $x \in \mathbf{X}$) to denote a matrix of unspecified dimension. 
Similarly, subscript $\mathbf{X}_{ij}$ denotes the $i$ number of rows and $j$ number of columns of the matrix $\mathbf{X}$. 
We use $\mathbb{E}[y|x]$ to express the expectations on variable $y$ given $x$, where the $|$ notation represents conditional likelihood. $P(x)$ or $p(x)$ are used for expressing the probability of $x$, we use them interchangeably without making a distinction on if the probability distribution is continuous or discrete. 
Finally, the falsum symbol, $\bot$, is dedicated to represent statistical independence throughout this paper.

Let us consider an experiment in which we introduce two software variants to two groups, control ($N_c$) and treatment ($N_t$). We denote the two levels of software variants as $t = \{0, 1\}$.
For each individual in the sample population $n \in N$, we measure a target variable, $y$, to understand the possible outcomes, and a set of covariates $x \in \mathbf{X}$ that are predictive of the outcomes and potentially influence the treatment. The covariates $\mathbf{X}$, are also often referred to as context, or confounding factors. 
In practise, the two treatment levels of software could be $t = \{\mathrm{old\_version}, \mathrm{new\_version}\}$, for evaluating a change to an existing software. The potential outcomes of such an evaluation of an energy management software, could be reported as $ y = [150, 300]$ measured in $\mathrm{Wh/km}$.

Suppose we are interested in two treatment levels in a study, $t = \{0, 1\}$. In the Rubin potential outcomes framework, the average treatment effect ($ATE$) can be expressed as,

\begin{equation}
    ATE = \mathbb{E}[y | t = 1] - \mathbb{E}[y | t = 0]
    \label{eq:ATE}
\end{equation}

where the $\mathbb{E}(y)$ represents the expectation of the outcome from the samples at different treatment levels.
In order to infer the treatment outcome, an important assumption made in the potential outcomes framework, is the stable unit treatment assumption, stating that the treatment only effect the individual sample the treatment is applied to.

Note that, when the experiment is randomised, the treatment outcome is unconditional to the control and treatment group assignment. In other words, the two groups are exchangeable. Therefore, we can explicitly establish a causal relationship between treatment and the potential outcome in a randomised experiment. Exchangeability is expressed as,

\begin{equation}
    [y|t=0], [y|t=1] \bot t
    \label{eq:exe}
\end{equation}

where $\bot$ denotes independence, and $|$ means given the condition, $y \bot t | x$ reads as $y$ is independent of $t$ given $x$.

The potential outcome and/or treatment assignment in an observational study is influenced by covariates $\mathbf{X}$. 
In a trivial example, all covariates influence both the treatment assignment and the outcome, a confounding effect.
We illustrate such an example in a directed acyclic graph (DAG) in Fig. \ref{figure:DAG}, to infer causality from this DAG, a valid adjustment set of covariates should block every path from the treatment $t$ to the target variable $y$ in the DAG. 
If such a set of covariates exists and can be identified in an observational study, we assume the exchangeability persists, and it is conditional on the adjustment set ($\mathbf{X}$) given the set of covariates can be observed and identified. Formally, 

\begin{equation}
    [y|t=0], [y|t=1] \bot t | \textbf{X}
    \label{eq:conditional_exe}
\end{equation}

In an observational study, it requires that there shall be treated and untreated samples in every combination of the values of the observed confounding factors $\mathbf{X}$ \cite{Westreich2010}, the positivity assumption, is formally expressed as,

\begin{equation}
    0 < P(T = t | \mathbf{X} = x|) < 1
\end{equation}

In addition, covariates can be used to estimate the conditional average treatment effect ($CATE$) in observational studies, namely,

\begin{equation}
    CATE = \mathbb{E}[y | t = 1, \mathbf{X} = x] - \mathbb{E}[y | t = 0, \mathbf{X} = x]
    \label{eq:CATE}
\end{equation}

The inference of conditional average treatment effect is helpful in heterogeneous studies. For example, we can study the treatment effect in subgroups of vehicle models or locations, provided the covariates heavily influence the treatment outcome.


\subsection{Bayesian statistics and inference}

A short overview of Bayesian statics and inference methods will be provided in this subsection. To put the Bayesian causal inference models in context, we present the the basic principle of Bayes' theorem and inference methods used in this study. 
However, we do not provide a comparison of frequency and Bayesian statistics, as the difference in reasoning is beyond the scope of this paper and we refer such a comparison to other works in software engineering \cite{Furia2019, Torkar2020, mattos2021statistical}.

In Bayesian statics, the probability of an event is expressed as a degree of belief which is based on the prior knowledge of said event. 
The intuition of Bayesian statistics is that the degree of belief is updated by observing new data, evidence, and the sensitivity of the outcome to the prior reduces as more observations are made.
In the application of causal inference, applying Bayesian statistics allows us to incorporate available prior knowledge on model parameters when inferring the counterfactual outcome \cite{Brodersen2015}.
The Bayes' theorem is expressed as,

\begin{equation}
    P(A|B) = \frac{P(B|A)P(A)}{P(B)}
    \label{eq:bayes}
\end{equation}

where,

\begin{itemize}
    \item $P(A)$ is the prior, the marginal probability of a hypothesis before any evidence is observed or presented. This is often referred to as domain knowledge.
    \item $P(B)$ is the marginal probability of observing the event $B$.
    \item $P(B|A)$: is the likelihood describing the probably of observing the evidence given the prior. 
    \item $P(A|B)$: is the posterior probability given the evidence, the observation, and the prior.
\end{itemize}

In most cases, the exact posterior distribution of the model parameters cannot be solved analytically, but it can be approximated numerically with Markov Chain Monte Carlo (MCMC) or variational inference methods. In this paper, we approximate the posterior distribution through the No-U-Turn Sampler (NUTS) in the Hamiltonian Monte Carlo algorithm. Using a recursive algorithm, NUTS constructs a set of possible candidate point spans widely across the target distribution \cite{Hoffman2011}. NUTS stops automatically if it retraced its steps, hence the name ``No-U-Turn''.

The prior distributions are an integral part of Bayesian model and they allow researchers the flexibility of incorporating domain knowledge of previous research to create better and more robust models. The priors often act as constraints of plausible probabilities of parameter values. With small sample sizes, the prior or the domain knowledge has a higher influence. While with larger sample sizes, the evidence overcomes the impact of the prior in the posterior. Priors can be specified to be \textit{non-informative}, \textit{weakly informative}, and \textit{informative}  for their models. A non-informative prior is based on an unbounded uniform distribution and does not aggregate any information to the posterior and are often non proper. Weakly-informative  are those that do not aggregate much information in the posterior parameters and serves as regularisers in the inference and convergence of the MCMC solver. An example of such a prior would be a normal distribution with a large variance compared to the expected parameter value. Finally, informative priors are those that incorporate domain knowledge on the subject and set stricter bounds to parameters in the model. If the evidence is accordance with the prior, convergence happens faster due to the smaller search space for the MCMC solver. If evidence points out to a parameter outside these bounds, either domain knowledge or data collection should be revised and convergence might be slow.

Comparing to another flavour of statistics, the frequentist statistics, Bayesian statistics has many benefits that have been discussion in previous literature.
By incorporating the concept of a prior, the stronger the prior information is, the less sample size Bayesian statistics requires to arrive a confident prediction of the posterior distribution. Utilising this property, Bayesian causal inference models applied in other areas of science, have reported to be less sensitive to sample sizes \cite{Chaplin2018, Geneletti2015, Normington2019, Li2020}. Moreover, instead of a point estimate prediction provided by the causal inference models in the frequentist domain, Bayesian causal inference models provide a complete posterior distribution, that can be utilised for the analysis of the treatment effect estimations. 





\section{The BOAT framework \label{theory}}

In this section, we present an alternative approach to randomised experimentation in software engineering, BOAT (\textbf{B}ayesian causal modelling for \textbf{O}bvserv\textbf{A}tional \textbf{T}esting). In this framework, we combine the notion of quasi-random treatment assignment with data obtained from pure observations, aiming to address the situations where a fully randomised experiment is not feasible.
Different from an observational study, in which the treatment is inferred from known systematic differences of the control and the treatment groups, our method allows one to actively intervene with a treatment group that is not randomly sampled. This framework enables development organisations to evaluate their software online without the need of a fully randomised large scale experiment.

\subsection{BOAT}

A fully randomised experiment is often challenging to conduct in the automotive domain, in the absence of randomisation, it requires a series of causal modelling techniques to mimic randomisation or to adjust covariates before a treatment effect can be inferred.
To address the challenges of randomisation, the BOAT framework is induced from the potential outcomes theory, and is applied and validated through exemplary cases with our industry collaborator. We describe the framework in detail as well as the research method applied for the validation of the BOAT framework.
We list the challenges in randomisation and their corresponding solutions in the BOAT framework as the following. 

The first challenge in adopting online experimentation is the limited access to the entire user base. 
To start, the automotive domain has a significantly smaller user based comparing to the SaaS domain, as a result of product diversity and hardware dependency \cite{Mattos2020, Liu2021a}. 
Moreover, in combination with the limitation of safety-critical software and the lack of explicit user agreements, shipping new software to the entire fleet is typically undesired, impossible, or unethical.
As a result, the control and the treatment groups are likely to be unbalanced and the treatment effects are confounded by one or more unbalanced covariates.
In situations when the treatment effect is confounded by more than one covariate, it is impossible to balance the control and treatment groups by stratification, therefore, a propensity score needs to be modelled from all covariates included in the system.
Propensity score matching, first proposed by Rosenbaum and Rubin \cite{Rosenbaum1983}, is a method for matching samples from the control and treatment group based on the propensity score calculated from observed covariates, thus adjusting for covariates and estimating unbias treatment effects. 
The Bayesian propensity score matching model is used for designing a balanced control and treatment group in traffic safety analysis \cite{Li2020} and in automotive software engineering \cite{Liu2021a}.

\begin{figure}[t]
\centerline{\includegraphics[width=\linewidth]{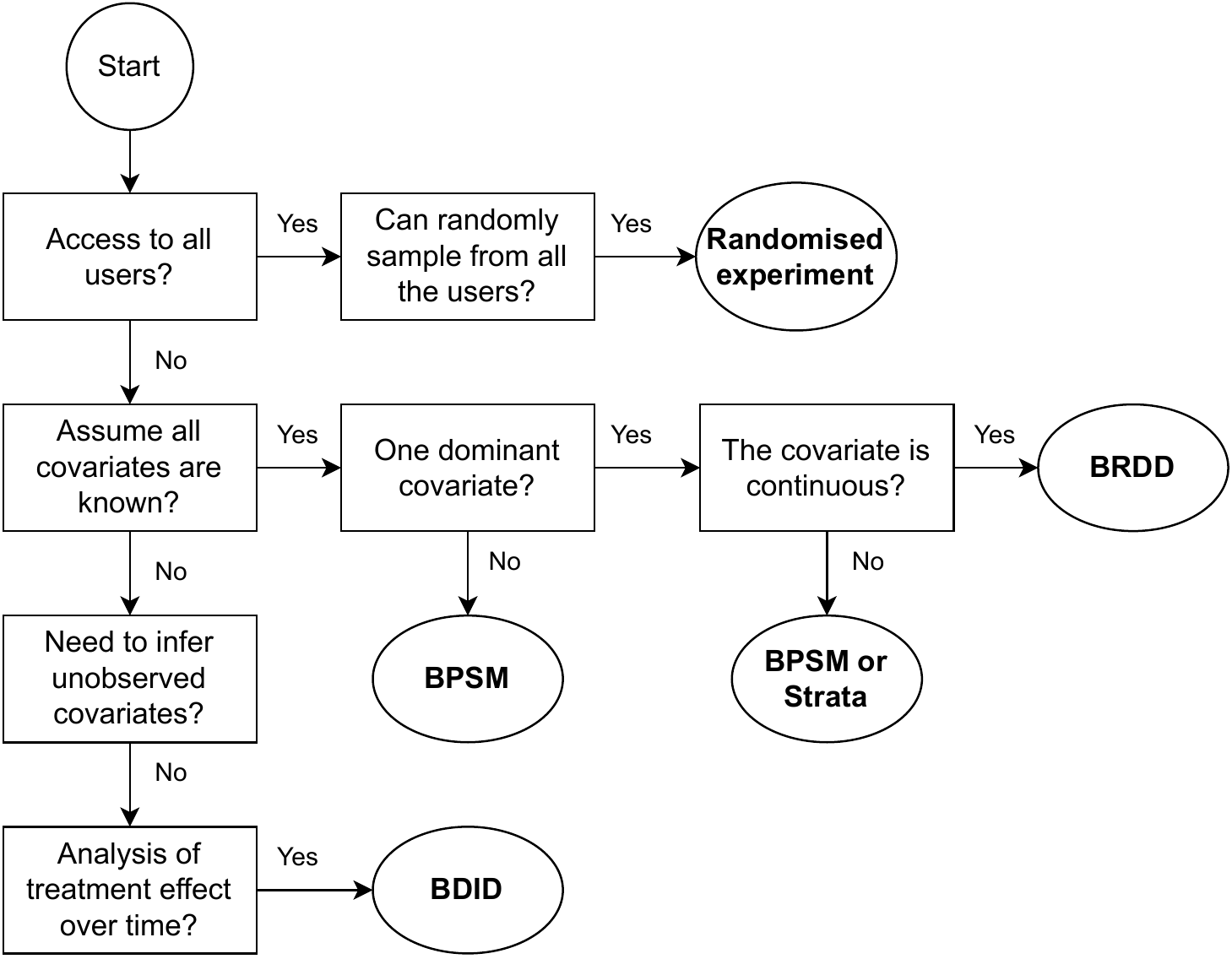}}
\caption{A decision flowchart on which Bayesian causal model from the BOAT framework to apply when designing an online software evaluation. (BRDD: Bayesian regression discontinuity, BPSM: Bayesian propensity score matching, BDID: Bayesian difference-in-differences)}
\label{fig_boat}
\end{figure}

Second, the performance of automotive software functions is often heavily influenced by temporal factors such as weather and time of week. Such a seasonality effect can be observed in software functions related to energy consumption \cite{Ahmed2017} and crash safety \cite{Bao2019, Kamel2021}.
In practise, if we want to evaluate an energy management software for battery electric vehicles, suppose we compare the energy efficiency before and after the software change on the same vehicles, the conclusion could be bias and confounded by unobserved temporal factors, e.g., temperature. 
To address this particular issue, we suggest to apply the difference-in-differences model in the Bayesian framework. 
This model is first presented by Card and Krueger \cite{Card1993} in analysing the treatment effects of increasing the minimum wage. 
In our model, we include a control group of vehicles running on the old software version through observation. Therefore, any bias caused by factors common to the control and treatment is implicitly controlled for, even when the confounders are not observed.
Besides econometrics, Bayesian difference-in-differences model is applied to analyse the treatment effect over time for diabetes patients \cite{Normington2019}. 

Third, we see a need of modelling and analysing software studies where the treatment assignments depend on one continuous covariate. 
Since many software functions in vehicles are only activated or beneficial for users around a certain threshold of a variable, such as speed or trip distance. 
This is the case when we are analysing the fuel saving potential from route prediction of plug-in hybrid vehicles, as the trip distance heavily influences the prediction accuracy and the fuel saving potential. It is believed that the software is particularly beneficial, when the driver travels slightly further than the pure electric range of the vehicle on the daily basis.
In this scenario, the regression discontinuity design \cite{Thistlethwaite1960} could help us model the treatment causal effect through identifying a threshold of an assignment covariate where the treatment is mostly influenced.
Bayesian regression discontinuity design is applied in other areas of science such as economics \cite{Chaplin2018} and medicine \cite{Geneletti2015}.

\begin{table*}[t]
\caption{Guidelines of design science research method \cite{Imbens1994}, and practices applied following the guidelines in this research.}
\centering
\begin{tabular}{|p{0.3\linewidth}|p{0.65\linewidth}|}
\hline
\multicolumn{1}{|c|}{\textbf{Guideline}} & \textbf{Practise in this research}      \\ \hline
Guideline 1: Design as an artifact       &  We present the BOAT framework to three software development organisations, the framework addresses varies limitations in randomised online experiments, assess practical scenarios to provide Bayesian causal modelling suggestions. The framework is derived from the theory of potential outcome, and all of the modelling approaches within the framework are extensively applied and validated in other areas of science \cite{Rosenbaum1983, Rubin2001, Xu2009, Li2020, Lee2008, Chib2015, Normington2019}.\\ \hline
Guideline 2: Problem relevance           &  To ensure the technical solution developed is relevant to the domain, we derive the problem from existing literature addressing the challenges on online experimentation adaptation in automotive \cite{Mattos2020, Liu2021, Liu2021a, Liu2021b,Giaimo2019}, and literature stating the challenges of online experimentation in other domains \cite{Deng2013, google2010, Xie2016}. All of the literature included in the analysis are based on empirical research in their respective domains. Additionally, the scenarios that limit randomised experimentation are also experienced and reported by our industry collaborator.                                       \\ \hline
Guideline 3: Design evaluation           & The evaluation of the BOAT framework is done quantitatively through three separate empirical cases. The empirical study are designed together with development organisations as suggested by \cite{Rubin2001, Xu2009, Stuart2010}, and deployed to a selective number of customer vehicles to simulate three scenarios of online software evaluation in the absent of randomisation. The quantitative data is collected from the vehicles through telecommunication units onboard. We determine the target variable and the covariates together with the development teams which also developed the software changes. Additionally, we assess the validity of the causal assumptions in practise with domain knowledge provided by experts from the development organisations.                                                   \\ \hline
Guideline 4: Research contributions      &  Not to be confused with the research contribution of a publication, the research contribution in design science is assessed by implementability and  representational fidelity. The former criteria is satisfied as there are commonly available tools for computing the the causal models, as well as the authors of this paper have implemented the code in Pyro \cite{bingham2019pyro}. The later is ensured through the close collaboration with an automotive manufacturer.                               \\ \hline
Guideline 5: Research rigor              &    Conducting research with design science often requires mathematical formalism to describe the design artifact \cite{Imbens1994, Peffers2007}. We formulate the Bayesian causal models and assumptions in the BOAT framework mathematically, as well as providing an algorithmic description for model implementation. In the BOAT framework, the relationships between factors within the system is assumed to be known; the inputs, such as covariates, and the outputs are selected based on the domain knowledge. Claims about the quality of design artifact is dependent on the choice of performance metrics. Therefore, we evaluate the causal models following advice from existing literature introducing the models to other areas of science \cite{Rosenbaum1983, Card1993, Thistlethwaite1960}.                                                         \\ \hline
Guideline 6: Design as a search process  & As a research method, design science is iterative and a way to discover an effective solution to the problem. In our research, we maintained close collaboration with the development teams through weekly design meetings, in which we discuss all aspects of the cases including the potential confounding factors, the expected treatment effects, and etc. The selection of covariates is done in an iterative manner to ensure the covariates with strong correlation to the target variable is included in the model. Moreover, the scenarios addressed in the BOAT framework is derived from literature \cite{Giaimo2020, Mattos2020} and validated from the state-of-practise in automotive software engineering.                                                                \\ \hline
Guideline 7: Communication of research   & The results of our research is communicated in the following three ways; (1) we host popularised science presentations for our industry collaborators on a regular basis every three month. The participants of these presentations usually occupy managerial positions such as product manager, project manager, and product owner. (2) we communicate the BOAT framework, the implemented Python code, and the results to the development organisations through our weekly meetings, during which we also provide tutorials of the Bayesian modelling approach. Participants are usually developers and data scientists alike. (3) we communicate our work to the scientific community through publications.               \\ \hline
\end{tabular}

\label{table_dsbs}
\end{table*}


To further illustrate the use cases of the BOAT framework in automotive software engineering, we provide a flow chart in Fig. \ref{fig_boat}. When designing a software online evaluation, the first assessment criteria is the available sample size determined by a power analysis of expected size of the treatment effect. 
In the scenario of all users can be accessed and a randomised sampling process can be done, a fully randomised experiment is always more ideal for a strong causal conclusion. Since randomisation for sampling from the total user base not only ensures exchangeability, it also ensures representativeness thus removing sampling bias. The second step is to determine if we can safely assume all the covariates and their relationship to the expected outcome is known and observable. 
Under this assumption, there is a need to identify if there are one or more covariates, since the balancing of a single and categorical covariate can be achieved through stratification as well. 
In situations where there are more than one categorical covariate, a Bayesian propensity score matching (BPSM) shall be performed to design a balanced control and treatment group. 
In the case of one continuous and dominating covariate, the design calls for the use of a Bayesian regression discontinuity (BRDD) model, which utilises the continuous covariate as an assignment variable for assigning samples to the control and treatment group.
If the assumption of known covariates cannot be made, which is often the case when evaluating a novel software functionality, or conducting a longitudinal evaluation. 
In the former scenario, there is usually no available data to analyse the causal structure, and in the latter case, the number of covariates required for identification quickly scales as the study is conducted during a prolonged period of time. 
The development organisations need to decide if the unobserved (latent) variables need to be modelled and inferred. If not, a Bayesian difference-in-differences (BDID) model can be applied. BDID is an effective strategy to infer treatment effect overtime, without the need of observing any time dependent covariates. 

Although the BOAT framework provides a structured guidance for software development organisations for their design decisions of online software evaluations, and to a large extend, enables such online evaluation that is otherwise challenging or impossible particularly in the automotive domain. We recognise that the framework is not in anyway complete. 
For example, if the causal relation is unknown, a causal discovery process \cite{Glymour2019} or a graphical model is needed \cite{Mattos2022}. 
To that end, if a latent variable is deemed to be important and needs to be modelled, methods such as instrumental variable \cite{Imbens1994} can be applied, as indicated as the missing "yes" action from latent variable inference in Fig. \ref{fig_boat}. We will further discuss the limitations and the potential extension of the BOAT framework in the discussion section.

\subsection{Research Method}




To validate our proposed BOAT framework and its applicability in automotive software engineering, we employ the design science research method following guidelines from \cite{Hevner2004}. 
The BOAT framework, can be considered as a design artifact created to address an important research and organisational problem -- in the absence of randomisation, how do we evaluate software in an online fashion and infer causality. The relevance of the problem is assessed and addressed through the challenges of randomisation experienced in practicse. 
We list the practices of this research following guidelines from \cite{Hevner2004} in detail in Table. \ref{table_dsbs}. In Table. \ref{table_dsbs}, we replace the original description from \cite{Hevner2004} of the step-wise guideline with the approaches taken in our research activities.
The industry collaborator is an automotive manufacturer with operations in Europe, China, and North America, and this research is part of a long term collaboration.
During this study, we worked closely with three software development teams, these teams are responsible for the conceptualisation, design, and development of their perspective software features in-house.

Design science is an iterative process containing three major cycles \cite{Marian2011}, one, the relevance cycle in which the problems and requirements are evaluated in the context. Two, the design cycle, during this step, the design artefacts and processes are proposed and evaluated. Three, the final design solution which has its base in scientific theory is then evaluated empirically.
In the first cycle of this research work, we aim to identify relevant challenges and limitation in the automotive domain for randomised experimentation. This research cycle is done in two ways. 
First, to ensure the relevance, we host weekly workshops and meetings with the software development teams to discuss and summarise the challenges experienced in practise. At least one of the researchers participate the meetings and workshops, and document the outcome as meeting notes.
To improve the external validity of the conclusions, we then map the challenges experienced in practise with reports from literature documenting online experimentation in the domain. 
Only challenges that are reported by two or more developers, or reported by a single developer and also reported in literature, are included in the final conclusion of the relevance cycle.

The design of the BOAT framework is a combination of addressing relevant challenges in randomisation, and solving them with the causal inference models introduced and applied in other domains. 
We study the theory of the causal inference models, as well as their applications and feasibility studies in other area of science through literature. 
We then explain the models with examples from other domains to the software development teams, both on the conceptual level and on the theoretical level. In this process, members of the software development teams provide detail description of possible use cases of the inference models. The descriptions are used as inputs for the following two aspects of the BOAT design, (1) helps us in evaluating which causal inference models are applicable in the automotive domain, (2) we summarise the most commonly mentioned use cases and design our empirical validation studies around them. 
Similarly to cycle one, this phase of the research is conducted through weekly meetings with the software development teams. At least one of the researchers participated in those meetings \textit{in situ}.

Last but not least, the successful adoption of a new framework in software engineering requires the framework to be investigated in the business organisations it is applied in \cite{Runeson2008}, therefore, together with our industry collaborator, we validate the framework empirically with three software development organisations. 
To evaluate the Bayesian causal models, we design three study cases, we deploy software and collect empirical data from a total of 1,364 vehicles driven by real-world customers. The vehicle data collection took place during an nine-month period, between December 2020 to September 2021. The detailed setup of the three empirical cases are discussed in their corresponding sections.



\section{Bayesian propensity score matching \label{BPSM}}

In this section, we present the theory, our observational study, and the results from the Bayesian propensity score matching (BPSM) model. 
The theory and assumption of the model is presented formally, followed by a discussion on different matching strategies. Finally, we introduce the setup of our observational study utilising BPSM as an identification strategy and we present the results.

\begin{figure}[t]
\centerline{\includegraphics[width=\linewidth]{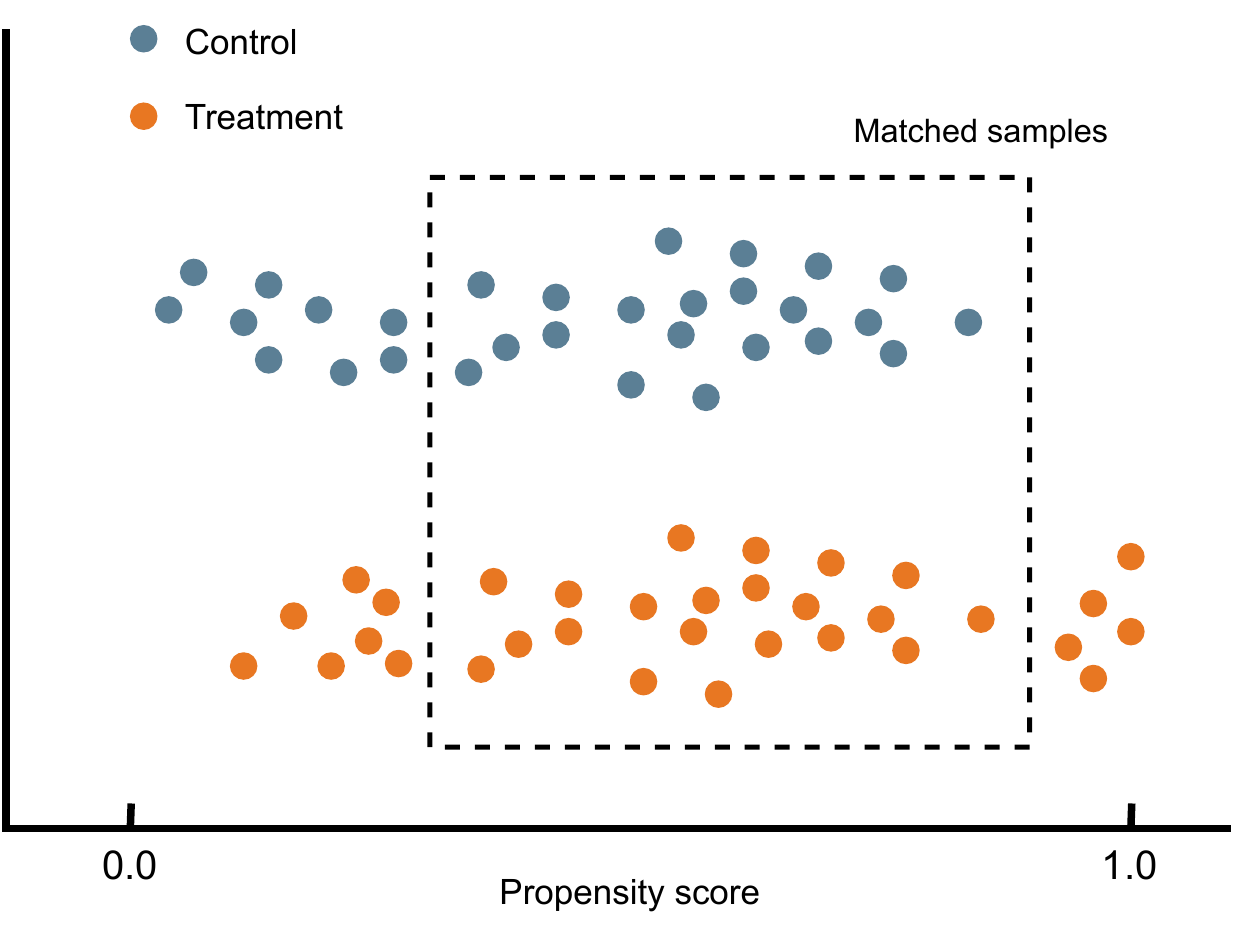}}
\caption{An illustration explaining the Propensity Score Matching model. Note, figure does not represent real data.}
\label{fig_PSM}
\end{figure}

To estimate the unbias treatment effect in an observational study is challenging, as the treatment assignment and effect could be confounded by covariates. 
We use the illustration in Fig. \ref{figure:DAG} to demonstrate this scenario.
Propensity score matching \cite{Rosenbaum1983} addresses this issue through covariate adjustment and allows us to generate balanced control and treatment groups even when the sample size is limited. The covariates from matched control and treatment group should form similar empirical distribution, thus reducing bias in the estimated treatment effect.
We illustrate the concept of propensity score matching in Fig. \ref{fig_PSM}, as can be seen, samples are matched based on their propensity score similarity, also called propensity score distance.
In previous literature, propensity score matching has been used for experiment design when the sample size is small \cite{Liu2021a, Rubin2001, Xu2009, Stuart2010a}, and for causal effect analysis \textit{post facto} \cite{Xu2016, Li2020}.
When propensity score estimate is done through a Bayesian network, literature reports the sensitivity to sample size being lower comparing to the conventional propensity score model \cite{Li2020}.

\subsection{Theory}

Propensity score, denoted as $e(x)$, is a function of one or more covariates $x \in \textbf{X}$. 
Propensity scores are modelled and matched in the control ($t=0$) and treatment ($t=1$) groups, so that the conditional probability distribution of the covariates given the propensity score $p(x|e(x))$ is similar in both groups \cite{Rosenbaum1983}.
The strong exchangeability assumption, extends from the conditional exchangeability, stating that the control and treatment outcome pair is assumed to be independent from treatment assignment, given the observed covariates, formally,

\begin{equation}
    ([y|t=0], [y|t=1]) \bot t | \textbf{X}
    \label{eq:strong_ignore}
\end{equation}

This assumption implies that the treatment or the outcome is only confounded by observed covariates and the covariates that are ignored from the data do not effect the treatment or outcome.
The average treatment effect identified through propensity score matching ($ATE_{PSM}$)  is conditional to the treatment $t$ and the propensity score inferred from all observed covariates, $e(\textbf{X})$. Given the assumption holds, the average treatment effect adjusted with the propensity score is an unbias estimate of the average treatment effect,

\begin{equation}
       ATE_{PSM} = \mathbb{E}[y | t = 1, e(\mathbf{X})] - \mathbb{E}[y | t = 0, e(\mathbf{X})]
\end{equation}

There are two steps in propensity score matching. First, we estimate the propensity score through a Bayesian logistic regression, then we perform the matching of samples based on their propensity score distance.
We present the two steps separately in the following subsections.

\subsubsection{Bayesian logistics regression}

In a BPSM, the propensity score is estimated with a Bayesian logistic regression.
Sticking to the same notations, the treatment indicator $t$, follows a Bernoulli distribution,

\begin{equation}
    t \sim Bernoulli(t|e(\mathbf{X}))
\end{equation}

where the propensity score $e(\mathbf{X})$ is expressed as,
\begin{equation}
    e(\mathbf{X}) = \frac{e^{\alpha  + \beta \mathbf{X}}}{1 + e^{\alpha  + \beta \mathbf{X}}}  
\label{eq:BPSM}
\end{equation}

The regression intercept and coefficients, $\alpha$ and $\beta$ are latent variables. 
That is, they are not directly observed but inferred from other variables $\mathbf{X}$. 
We normalise the prior Gaussian distributions for the regression intercept $\alpha$, as a result, this prior distribution has a $0$ mean and a variance of $\lambda_{\alpha}$,
\begin{equation*}
  \alpha \sim \mathcal{N}(\alpha|0, \lambda_{\alpha})
\end{equation*}

similarly, $\beta$ has a Gaussian distributions of a $0$ mean and variance of $\lambda_{\beta}$ as prior,
\begin{equation*}
    \beta \sim \mathcal{N}(\beta|0, \lambda_{\beta})
\end{equation*}

By Bay's theorem, the posterior distribution of this network is simply the product of the likelihood and the prior.
Therefore, for all samples $n \in N$, the posterior distribution is a joint probability of $t$, $\alpha$, and $\beta$ marginalised over $p(t)$, that is,

\begin{equation}
\begin{aligned}
    &p(t, \alpha, \beta | \mathbf{X}, \lambda_{\alpha}, \lambda_{\beta}) \\
        &= p(\alpha|\lambda_{\alpha}) \cdot  p(\beta|\lambda_{\beta}) \cdot \prod_{n=1}^{N}p(t|\alpha, \beta, \mathbf{X})
\end{aligned}
\end{equation}

Bayesian networks are generative models, to generate the joint probability distribution of the regression model, the generative process is stated in Algorithm \ref{bpsm_gen}.

\begin{algorithm}[t]
\caption{Bayesian logistic regression generative process}
\textbf{Inputs:} $\mathbf{X}$ covariates, $\lambda_{\alpha}$ prior distribution of $\alpha$, $\lambda_{\beta}$ prior distribution of $\beta$, $t_n$ control/treatment indicator
\begin{algorithmic}[1]
\State Draw $\alpha \sim \mathcal{N}(\alpha|0, \lambda_{\alpha})$
\State Draw $\beta \sim \mathcal{N}(\beta|0, \lambda_{\beta})$

\For{each vector of covariate $x \in \mathbf{X}$}
     \State Draw $t \sim Bernoulli(t|Sigmoid(\alpha + \beta x ))$
\EndFor
\end{algorithmic}
\label{bpsm_gen}
\end{algorithm}

    \subsubsection{Matching}

The second and final step of BPSM is matching samples from the control and treatment groups based on their propensity score distance, and to minimise the average distance for the two groups.
The propensity score distance ($\delta e(\mathbf{X})_n$) is defined as the absolute difference of the propensity score of each sample ($n$) in the control and treatment group,

\begin{equation}
    \delta e(\mathbf{X})_{n} = |e(\mathbf{X})_{n, t=1} - e(\mathbf{X})_{n, t=0}|
\end{equation}

Many matching methods have been explored in the literature, calliper matching \cite{Austin2011}, 1:1, or n:1 nearest neighbour matching \cite{Stuart2010a}, and full matching \cite{Xu2009, Hansen2004}, just to name a few.
In general, there are two categories of matching methods, with or without replacement.
Matching can be done with or without replacement.
Matching with replacement means one sample in one group can be matched with multiple samples in another group, an example of such method is the optimal full matching algorithm \cite{Hansen2004}.

\begin{table*}[ht]
\caption{Descriptive statistics of the target variable and covariates as inputs to Bayesian propensity score matching model, and a description of how the variables are computed. Each variable is aggregated to the vehicle level and min-max scaled.}
\centering
\begin{tabular}{lllll}
\hline
\textbf{Variables} & \textbf{Variable description} & \textbf{Group} & \textbf{Mean} & \textbf{Std.} \\ \hline\hline
\textbf{Target variable} &  &  &  &  \\
\multirow{2}{*}{Fuel consumption {[}g/km{]}} & \multirow{2}{*}{total fuel injected in engine / total distance} & Control & 0.391 & 0.155 \\
 &  & Treatment & 0.354 & 0.123 \\ \hline
\textbf{Covariates} &  &  &  &  \\
\multirow{2}{*}{Share of trip start at a high state-of-charge} & \multirow{2}{*}{number of trip where soc\_start \textgreater 80\%} & Control & 0.356 & 0.177 \\
 &  & Treatment & 0.397 & 0.178 \\
\multirow{2}{*}{Share of trip end at a low state-of-charge} & \multirow{2}{*}{number of trip where soc\_end \textless 21\%} & Control & 0.258 & 0.155 \\
 &  & Treatment & 0.120 & 0.150 \\
\multirow{2}{*}{Number of trips made on weekdays} & \multirow{2}{*}{number of trips taken place during weekdays} & Control & 0.290 & 0.167 \\
 &  & Treatment & 0.225 & 0.153 \\
\multirow{2}{*}{Number of trips made on weekends} & \multirow{2}{*}{number of trips taken place during weekends} & Control & 0.269 & 0.173 \\
 &  & Treatment & 0.190 & 0.154 \\
\multirow{2}{*}{Average trip distance {[}km{]}} & \multirow{2}{*}{total trip distance / total number of trips} & Control & 0.301 & 0.132 \\
 &  & Treatment & 0.343 & 0.124 \\
\multirow{2}{*}{Maximum trip distance {[}km{]}} & \multirow{2}{*}{longest trip occurred during the observation period} & Control & 0.278 & 0.193 \\
 &  & Treatment & 0.240 & 0.186 \\
\multirow{2}{*}{Average trip speed {[}km/h{]}} & \multirow{2}{*}{total trip distance / total trip duration} & Control & 0.575 & 0.110 \\
 &  & Treatment & 0.624 & 0.117 \\
\multirow{2}{*}{Maximum trip speed {[}km/h{]}} & \multirow{2}{*}{highest trip speed occurred during the observation period} & Control & 0.637 & 0.125 \\
 &  & Treatment & 0.640 & 0.135 \\
\multirow{2}{*}{Share of distance on "hybrid"} & \multirow{2}{*}{distance driven when vehicle is in hybrid mode / total distance} & Control & 0.956 & 0.103 \\
 &  & Treatment & 0.987 & 0.033 \\
\multirow{2}{*}{Share of trips with a trailer attached} & \multirow{2}{*}{numbers of trip with trailer attached / total number of trips} & Control & 0.034 & 0.081 \\
 &  & Treatment & 0.034 & 0.075 \\
\multirow{2}{*}{Average number of engine starts in a trip} & \multirow{2}{*}{total occurrence of engine RPM \textgreater 500 / total number of trips} & Control & 0.169 & 0.112 \\
 &  & Treatment & 0.176 & 0.176 \\
\multirow{2}{*}{Average ambient temperature {[}$^\circ C${]}} & \multirow{2}{*}{average temperature measured at car during the period} & Control & 0.371 & 0.084 \\
 &  & Treatment & 0.372 & 0.071 \\
\multirow{2}{*}{Minimum ambient temperature {[}$^\circ C${]}} & \multirow{2}{*}{minimum temperature measure at car during the period} & Control & 0.497 & 0.135 \\
 &  & Treatment & 0.563 & 0.150 \\
\multirow{2}{*}{Maximum ambient temperature {[}$^\circ C${]}} & \multirow{2}{*}{maximum temperature measure at car during the period} & Control & 0.388 & 0.101 \\
 &  & Treatment & 0.374 & 0.099 \\ \hline
\end{tabular}
\label{table_data_bpsm}
\end{table*}

By using matching methods without sample replacement, the matched control and treatment group will yield the number of samples.
Calliper matching is a type of matching method, in which a maximum allowed $\delta e(\mathbf{X})_{n, max}$ is predetermined and all samples exceeding the threshold are discarded. Although calliper matching could result in a reduction on sample size if the calliper is too fine, it is an intuitive and computationally efficient matching method \cite{Austin2011}.
Moreover, the choice of calliper can affect the result bias \cite{Wang2020}.
As treated samples are usually more expensive to obtain, discarding those samples could be considered unfavourable in the automotive application.
Therefore, a second matching method is explored in the study, 1:1 nearest neighbour matching \cite{Stuart2010a}.
In a 1:1 nearest neighbour matching, the algorithm looks for a control sample with the closest propensity score distance to a given treated sample, thus, no treated samples will be discarded.
In this study, both calliper matching and 1:1 nearest neighbour matching are applied, as our objective is to find control samples to be compared with the treated samples.

\begin{figure*}[ht]
\centerline{\includegraphics[width=\textwidth]{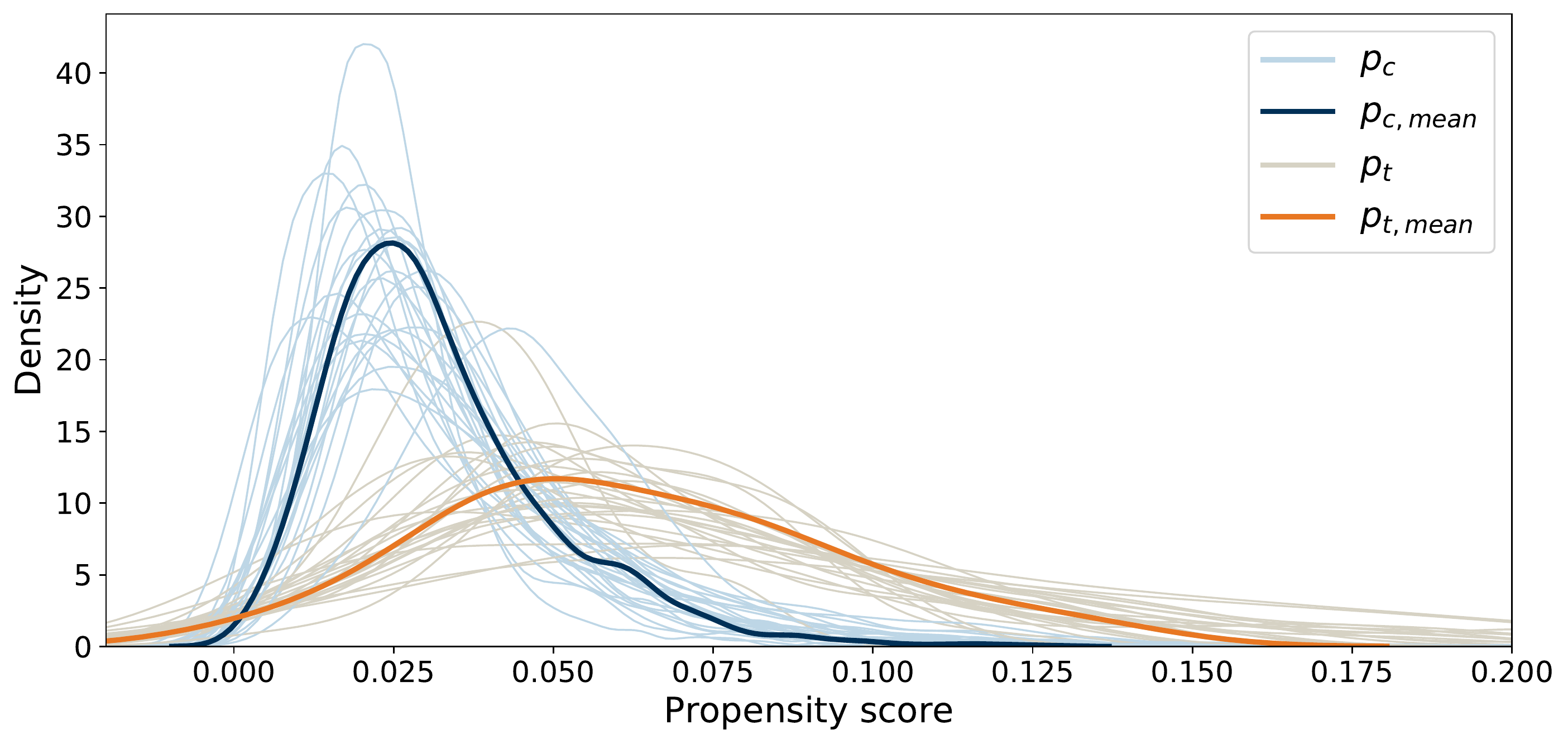}}
\caption{Kernel density distribution of the propensity scores of the control ($p_c$) and treatment ($p_t$) groups calculated on the mean of posterior distributions, and twenty-five values randomly sampled from the posterior distributions representing uncertainties.}
\label{fig_bpsm}
\end{figure*}

\subsection{Study I: Limited access to users}

In the automotive domain, a fully randomised experiment on a large scale is often impossible due to their already limited numbers of users and it is often undesired to serve new software on the entire fleet.
When a safety critical software is the subject of interest -- a majority of automotive software are in safety critical systems -- introducing a novel software feature to a larger number of vehicles might not be desirable; although the likelihood of catastrophic failure is low \cite{Mattos2020}, but any minor disturbances at a scale could still lead to financial loss for commercial vehicles. 
Combine this with the fact that we often only want to examine software features on a specific vehicle model driven in a specific region, it further limits the available sample size. 
As an alternative to large scale randomised experiments, we propose a small-scale rollout to a limited number of vehicles. However, when the control and treatment groups are not randomised, the exchangeability assumption does not hold and the observed change in the target variable could be confounded on preexisting differences between the groups instead of the treatment itself. 
Thus, we apply Bayesian propensity score matching for matching comparable treated and untreated vehicles based on a number of observed covariates.

We design a study to simulate this scenario and learn the feasibility of applying BPSM as an identification strategy in observational studies such as this.
In Study I, we aim to analyse treatment effects from an energy management software. Based on how the vehicle is historically driven, this software predicts the current trip proprieties such as expected route, and based on the prediction, energy consumption is optimised through actions such as downshifting before a hill climb.
This software was difficult, if not impossible, to validate in a lab-like environment as the energy saving potential is strongly dependent on the prediction accuracy, however, since the software feature involves safety critical systems such as engine control, it is undesirable to introduce it to a large group of users for a fully randomised experiment.
Instead, we passively observe 1100 vehicles in the control group running on an existing software driven by real-world customers, while only serving the modified software to 38 vehicles as our treated samples. The treated samples, are vehicles leased to employees whom use the vehicles as their regular cars. This study occurred from October 2020 to March 2021, and all vehicles are driven by users residing in Sweden. We discard data generated by brand new vehicles with mileage less than 100 kilometres. We also discard trip data in which the average trip speed is higher than 200 kilometres per hour. After postprocessing, we have collected data from a total of 421,881 trips made by 1138 vehicles.

A fundamental assumption of applying propensity score as an identification strategy for observational study is the strong exchangeability assumption (often called ignorability \cite{Rosenbaum1983}), implies that the treatment outcome is not dependent on unobserved covariates. 
This strong assumption cannot be checked empirically from pure observational data. 
Essentially, the assumption implies that all the confounders that potentially influence the outcome are observed and it inherently limits us to draw causal conclusion when no covariate is known. In other words, propensity score matching can only balance covariates that are observed, while full randomisation can balance all covariates, observed or not \cite{Rubin2001}.
In practice, to design a study utilising BPSM requires existing data or knowledge of the software systems, as the results are strongly dependent on covariate selection \cite{Brookhart2006}. To this end, an iterative procedure can be applied when selecting covariate and performing BPSM \cite{Rosenbaum1984}. In our study, we include a total of 14 covariates in the final BPSM model as a outcome of domain knowledge provided by our industry collaborator and two iterations of model design. We present the descriptive statistics of the covariates included in the final model in Table. \ref{table_data_bpsm}. 



\subsection{Results}

In this subsection, we present the results from Study I. First, we show the propensity score inferred from a Bayesian network. Second, we present matched control and treatment group with the two matching strategies discussed in the previous section.

We implement a Bayesian logistic regression in Pyro \cite{bingham2019pyro} following the generative process described in Algorithm \ref{bpsm_gen}.
We set up a NUTS sampler for inference to infer the posterior distribution with a single chain. We generate 3,000 samples of which 200 burn-in.
The Brooks-Gelman-Rubin convergence criteria of $\hat{R} < 1.1$ is met, at $\hat{R} = 1.0003$.
To triangulate the results, a variational inference model is used. We use a multivariate normal distribution as a guide for the variational inference model. We define 40,000 steps for optimisation and the solver reaches a stable solution after the first 10,000 steps.
Two inference methods return similar posterior distributions and point estimates.
We show both methods in the online appendix attached and we will only focus on reporting the inference results from the NUTS sampler in this paper.

We assign a prior distribution $\beta \sim \mathcal{N}(0, \lambda_{\alpha} = 1)$ to each regression coefficient $\beta$, and similarly, $\alpha \sim \mathcal{N}(0, \lambda_{\beta} = 1)$ to the regression intercept $\alpha$.
Combining the priors, the posterior, $p(t, \alpha, \beta | \mathbf{X}, \lambda_{\alpha}, \lambda_{\beta})$, is inferred.
Then, the propensity score $e(\mathbf{X})$ is calculated following Equation. \ref{eq:BPSM}, before matching, the mean propensity score in the control and treatment group is 0.0319 and 0.0633 respectively. 
In each group, the standard deviation of the propensity score is 0.0175 and 0.0309.
We also quantify the uncertainty of the propensity score from the posterior distribution.
A visualisation of the propensity score distribution before a matching is performed can be seen in Fig. \ref{fig_bpsm}, in which we plot the propensity score distribution in the control ($p_c$) and treatment ($p_t$) computed from the mean point estimates as well as random samples from the posterior distribution of the regression intercept and coefficients.

\begin{table}[t]
\caption{Propensity scores in control and treatment groups, before and after a matching is performed.}
\centering
\begin{tabular}{llll}\hline
Propensity score &  &  &  \\ 
 & Group & Mean & Std. \\ \hline\hline
\multirow{2}{*}{Before matching} & Control & 0.0319 & 0.0175 \\
 & Treatment & 0.0633 & 0.0309 \\
\multirow{2}{*}{Calliper matching (calliper = 0.05)} & Control & 0.0626 & 0.0300 \\
 & Treatment & 0.0633 & 0.0309 \\
\multirow{2}{*}{1-1 nearest neighbour matching} & Control & 0.0627 & 0.0302 \\
 & Treatment & 0.0633 & 0.0309 \\ \hline
\end{tabular}
\label{tabel_scores}
\end{table}

A matching based on propensity score distance $\delta e(\mathbf{X})_{n}$ is performed after the scores are computed. In this paper, we perform matching with 1:1 nearest neighbour and caliper matching method, the results are presented in Table \ref{tabel_scores}.
As can be seen from the table, both matching methods return similar outcome in this study, the mean propensity score distance is 0.000757 and 0.000608 for the calliper matching and 1-1 nearest neighbour matching respectively.
First, a calliper matching method is used with a specified maximum propensity score distance, $\delta e(\mathbf{X})_{n, max} = 0.05$, and every treated sample returned a matched control sample. 
The mean propensity scores in the control group were 0.0626 after applying caliper matching.

Second, we apply 1-1 nearest neighbour matching method.
Similarly to calliper matching, 1-1 nearest neighbour match will return one-to-one matched pairs, but the matching is done without specifying maximum allowed propensity score distance.
The algorithm k-nearest neighbours searches for one closest neighbour from the control samples to the treated samples. 
The mean propensity score in the control group is 0.0627 after 1-1 nearest neighbour matching.
Both methods find the corresponding control samples for the treated samples.
The covariates balance is assessed by comparing the descriptive statistics such as variance and the empirical distribution of covariates in the two groups.
With a calliper matching, we found an average of 4.1 \% reduction in the covariates variance compared to unmatched groups.

The treatment effect is analysed for both the calliper matched and 1-1 nearest neighbour matched groups.
The average treatment effect is calculated as the mean difference of the target variable between the control and treatment groups, and all values are min-max scaled.
The average target variable is 0.379 and 0.391 for the control group when matched with calliper and 1-1 nearest neighbour, respectively.  
The average target variable is 0.355 in the treatment group.
The average treatment effect is -0.024 and -0.036 for the control group when matched with calliper and 1-1 nearest neighbour, respectively.

\section{Bayesian difference-in-differences \label{BDID}}

\begin{figure}[t]
\centerline{\includegraphics[width=\linewidth]{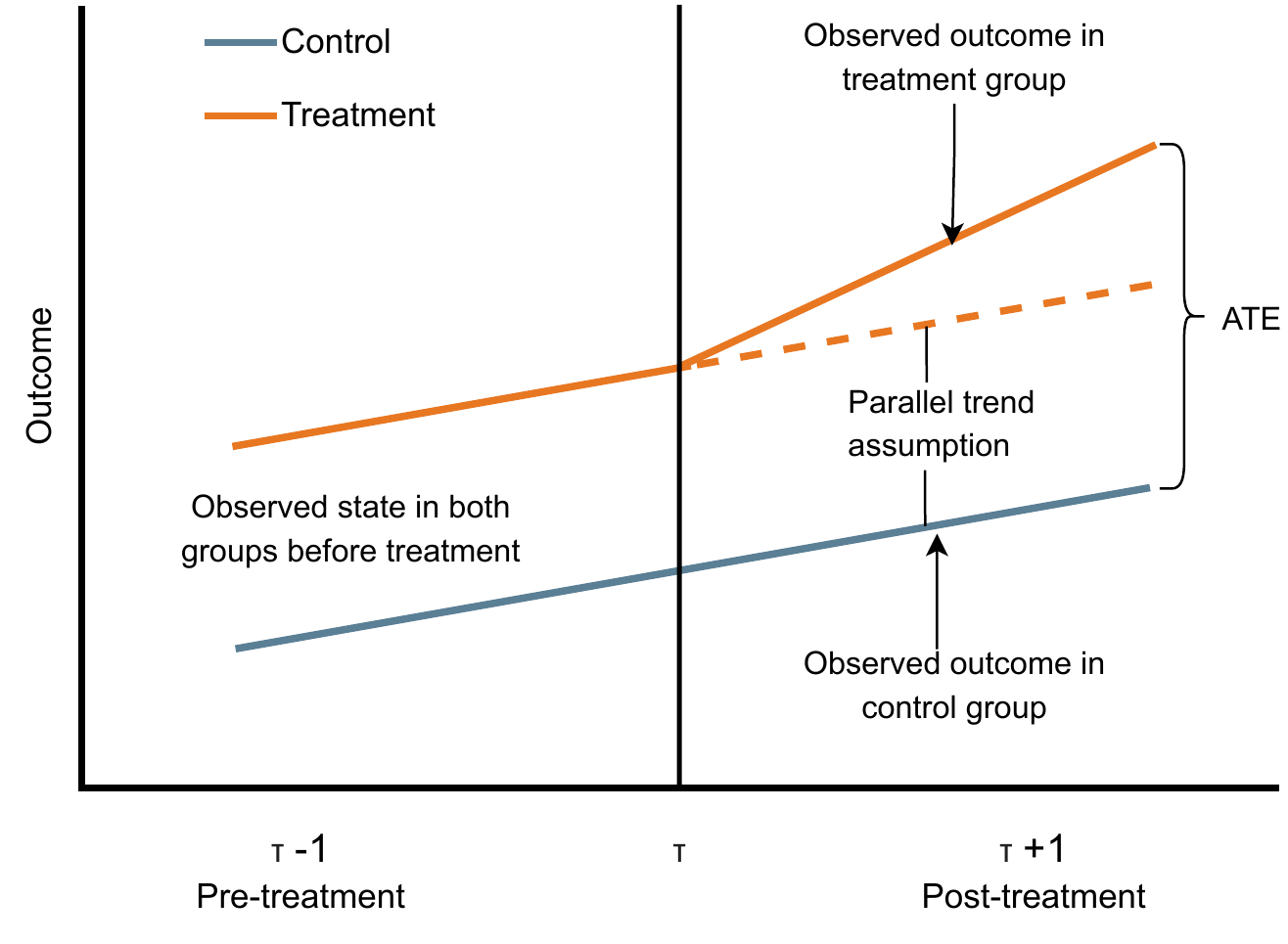}}
\caption{An illustration explaining the Difference-in-Differences model and how the average treatment effect (ATE) is estimated.}
\label{fig_DID}
\end{figure}

In this section, Bayesian difference-in-differences (BDID) theory and our observational study is presented.
We describe the theory and assumptions in the model formally, we present our study setup and how the model is utilised as an identification strategy for analysing treatment effects over time.

Difference-in-differences, proposed by Card and Krueger \cite{Card1993}, is a discrete time dynamic causal model. The model is designed to identify and control time-dependent covariates in observational studies, disregarding if the covariates are measured or not \cite{Normington2019}, and model the average treatment effect.
Different from cross-sectional treatment effect estimates, where the treatment effect is aggregated over time, or time-series treatment effect estimates, where time is treated as a continuous variable. This model can be used to measure treatment effects in discrete time steps.
We demonstrate the concept of the model in Fig. \ref{fig_DID}, as shown in the figure, the model replies on the parallel trend assumption, which implies the treatment group and control group are assumed to follow a similar trend for the target variable without a treatment.
The factual outcome is the observed target variable after the treatment is applied, the counterfactual outcome, what would have happened if a treatment is never applied, is inferred from the parallel trend assumption. 
Bayesian difference-in-differences have been applied in studying the influence of policy change on diabetes treatment quality \cite{Normington2019}. But there is no documented application of BDID in software engineering to the best of our knowledge.

\subsection{Theory}

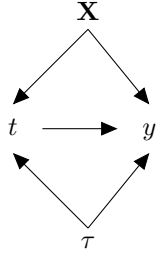
\begin{figure}[t]
\centering
\begin{tikzpicture}[
roundnode/.style={circle, draw=black!60, fill=white, very thick, minimum size=10mm},
roundedrect/.style={rectangle, rounded corners, minimum width=3cm, minimum height=1.5cm,text centered, draw=black}
]
\node      (maintopic)          {\begin{tabular}{c}$t$\end{tabular}};
\node      (uppercircle)       [above=of maintopic, xshift=1.0cm] {$\mathbf{X}$};
\node      (rightcircle)       [right=of maintopic] {\begin{tabular}{c} $y$ \end{tabular}};
\node      (latenttime)        [below=of maintopic, xshift=1.0cm] {$\tau$};

\draw[->] (uppercircle.south) -- (rightcircle.north);
\draw[->] (uppercircle.south) -- (maintopic.north);
\draw[->] (maintopic.east) -- (rightcircle.west);
\draw[->] (latenttime.north) -- (maintopic.south);
\draw[->] (latenttime.north) -- (rightcircle.south);

\end{tikzpicture}
\caption{A simplified directed acyclic graph showing the relationships of treatment ($t$), target variable ($y$), covariates ($\mathbf{X}$), and time dependent latent variables summarised as $\tau$.}

\label{figure:DAG_DID}
\vspace{-0.5cm}
\end{figure}

By including a group of untreated samples through passive observations from the same time period as the treated samples, all time dependent covariates are implicitly controlled for in a DID model, observed or not.
Recall the directed acyclic graph in Fig. \ref{figure:DAG}, we now extend it to include a $\tau$ variable to represent time dependent latent variables (Fig. \ref{figure:DAG_DID}), to conclude causal effect, both the covariates $\mathbf{X}$ and the time dependent latent variables $\tau$ need to be adjusted for.
In the difference-in-differences model, the most important assumption is the parallel trend assumption. It implies the counterfactual - what would have happened in the absent of a treatment - is an assumption inherently unobserved.
This assumption supports the exchangeability assumption as stated in Eq. \ref{eq:exe}, i.e., the treatment assignment is not based on the outcome, rather that the outcome is influenced by the applied treatment. 
We can express this assumption formally, note that we adopt the same notations from Section \ref{theory} and \ref{BPSM}. Additionally, we use $\tau = \{-1, 0, 1 \}$ to denote the time periods before, during, and after a treatment is applied.

\begin{equation}
       \mathbb{E}[y^0(1) - y^0(-1) | t = 1] = \mathbb{E}[y^0(1) - y^0(-1) | t = 0]
    \label{eq:parallel}
\end{equation}

In Eq. \ref{eq:parallel}, $y^{0}(-1)$ is the target variable with treatment level $0$ at time step $\tau = -1$ (pre-treatment status in Fig. \ref{BDID}), and $y^{0}(1)$ is the target variable with treatment level $0$ at time step $\tau = 1$ without a treatment being applied.
We use the superscript to represent the counterfactual status of the treatment group, if a treatment is never applied.
The parallel trend assumption states that the target variable measured from the control and treatment group will follow similar trend over time, if no treatment is applied at $\tau$, this is often referred as the counterfactual outcome. The parallel trend assumption can be check from observational data through means such as data visualisation.

The average treatment effect identified through difference-in-differences ($ATE_{DID}$) is estimated as the following,

\begin{equation}
\begin{aligned}
    &ATE_{DID} \\
    &= (\mathbb{E}[y(1) | t = 1] - \mathbb{E}[y(-1) | t = 1]) - \\
    &\ \ \ \ (\mathbb{E}[y(1) | t = 0] - \mathbb{E}[y(-1) | t = 0])
    \label{eq:ATE_DID}
\end{aligned}
\end{equation}

That is, the difference of the target variables in the treated group measured at time step $\tau = -1$ and $\tau = 1$, subtracted with the difference in the control group measured during the same time period, namely, the difference in differences. 
The target variable $y$ can be estimated as a linear regression from the data observed,

\begin{equation}
    y \sim t + \tau + \alpha + \beta \mathbf{X} + \epsilon
\end{equation}

In the regression model, we also assign a dummy variable $t$ indicating if a treatment is applied.
The latent variables regression intercept $\alpha$ and coefficient $\beta$ have a Gaussian distribution as prior, formally,

\begin{equation}
    \alpha \sim \mathcal{N}(\alpha|0, \lambda_{\alpha}) 
\end{equation}

Let us consider a total $j$ numbers of covariates $\mathbf{X}$, and regression coefficient is a vector of length $j$,

\begin{algorithm}[t]
\caption{Bayesian difference-in-differences generative process}
\textbf{Inputs:} $\mathbf{X}$ covariates, $\lambda_{\alpha}$ prior distribution of $\alpha$, $\lambda_{\beta}$ prior distribution of $\beta$, treatment effect $y_n$ 
\begin{algorithmic}[1]
\State Draw $\alpha \sim \mathcal{N}(\alpha|0, \lambda_{\alpha})$
\State Draw $\beta \sim \mathcal{N}(\beta|0, \lambda_{\beta})$
\State Draw $\epsilon \sim \mathcal{N}(\epsilon|0, \sigma^2)$

\For{each vector of covariate $x \in \mathbf{X}$}
     \State Draw $y \sim \mathcal{N}(y|t + \tau + \alpha +\beta^Tx, \sigma^2)$
\EndFor
\end{algorithmic}
\label{bdid_gen}
\end{algorithm}

\begin{equation}
    \beta_j \sim \mathcal{N}(\beta_j|0, \lambda_{\beta_j})
\end{equation}

Moreover, since we cannot describe all variations in the data with a linear model, a error term $\epsilon$ is included in the model which represent the observation noise. We have,

\begin{equation}
    \epsilon \sim \mathcal{N}(\epsilon | 0, \sigma^2)
\end{equation}

The joint distribution is factorised as the equation below, it is a straight forward application of Bay's theorem. Moreover, we list the generative process for this join distribution in Algorithm \ref{bdid_gen}.

\begin{table*}[t]
\caption{Descriptive statistics of the target variable and covariates as inputs to Bayesian difference-in-differences model, and a description of how the variables are computed. Each variable is aggregated to the vehicle level and min-max scaled.}
\centering
\begin{tabular}{lllll}
\hline
\textbf{Variables} & \textbf{Variable description} & \textbf{Group} & \textbf{Mean ($\tau_{-1}$)} & \textbf{Mean ($\tau_1$)} \\ \hline\hline
\textbf{Target variable} &  &  &  &  \\
\multirow{2}{*}{Energy consumption {[}Wh/km{]}} & \multirow{2}{*}{total electrical energy consumed / total distance} & Control & 0.277 & 0.296 \\
 &  & Treatment & 0.257 & 0.276 \\ \hline
\textbf{Covariates} &  &  &  &  \\
\multirow{2}{*}{Time period} & \multirow{2}{*}{dummy, 0 for pre-treatment and 1 otherwise} & Control &  &  \\
 &  & Treatment &  &  \\
 \multirow{2}{*}{Treatment} & \multirow{2}{*}{dummy, 0 for control group and 1 for treated group} & Control &  &  \\
 &  & Treatment &  &  \\
 \multirow{2}{*}{Average ambient temperature {[}$^\circ C${]}} & \multirow{2}{*}{average temperature measured at car} & Control & 0.667 & 0.318 \\
 &  & Treatment & 0.954 & 0.465 \\
\multirow{2}{*}{Minimum ambient temperature {[}$^\circ C${]}} & \multirow{2}{*}{minimum temperature measure at car} & Control & 0.630 & 0.294 \\
 &  & Treatment & 0.557 & 0.272 \\
\multirow{2}{*}{Maximum ambient temperature {[}$^\circ C${]}} & \multirow{2}{*}{maximum temperature measure at car} & Control & 0.429 & 0.484 \\
 &  & Treatment & 0.745 & 0.626 \\ 
\multirow{2}{*}{Average trip distance {[}km{]}} & \multirow{2}{*}{total trip distance / total number of trips} & Control & 0.257 & 0.192 \\
 &  & Treatment & 0.314 & 0.310 \\
\multirow{2}{*}{Maximum trip distance {[}km{]}} & \multirow{2}{*}{longest trip occurred during the observation} & Control & 0.229 & 0.322 \\
 &  & Treatment & 0.270 & 0.392 \\
\multirow{2}{*}{Average coolant temperature {[}$^\circ C${]}} & \multirow{2}{*}{coolant temperature measured at battery outlet} & Control & 0.559 & 0.333 \\
 &  & Treatment & 0.821 & 0.396 \\
\multirow{2}{*}{Average voltage battery discharge {[}Wh{]}} & \multirow{2}{*}{average battery energy discharge / number of trips} & Control & 0.612 & 0.587 \\
 &  & Treatment & 0.506 & 0.582 \\
\multirow{2}{*}{Average starting battery capacity {[}Wh{]}} & \multirow{2}{*}{average battery capacity measured at start of a trip} & Control & 0.635 & 0.608 \\
 &  & Treatment & 0.531 & 0.616 \\
 \multirow{2}{*}{Average state-of-charge change {[}\%{]}} & \multirow{2}{*}{average displacement of battery state-of-charge} & Control & 0.609 & 0.579 \\
 &  & Treatment & 0.484 & 0.562 \\ \hline
\end{tabular}
\label{table_data_bdid}
\end{table*}

\begin{equation}
\begin{aligned}
    &p(y, \alpha, \beta | \mathbf{X}, \sigma, \lambda_{\alpha}, \lambda_{\beta}) \\
    &= p(t) \cdot p(\alpha|\lambda_{\alpha}) \cdot  p(\beta_j|\lambda_{\beta_j}) \cdot \prod_{n=1}^{N}p(y|t, \alpha, \beta_j, \sigma, \mathbf{X})
\end{aligned}
\end{equation}

\subsection{Study II: Seasonality effect}

A large portion of software in the automotive domain is influenced by the vehicle operating conditions, e.g., precipitation, temperature, humidity, and icing \cite{Ahmed2017, Bao2019, Kamel2021}, and most importantly, the mobility needs of people.
In the first case, these operating conditions are often seasonal and from the diversity of vehicle markets today, they are difficult to predict beforehand and often requires the software to be evaluated longitudinally to cover a wider range of operating conditions and increase confidence in the conclusion.
Additionally, it is naturally reasonable to assume that there are seasonality effects that cannot be observed in an effective manner nor can it be predicted, such as public events, extreme weathers and so on.
The travel demand of users can be largely unpredictable similar to the operating conditions of the vehicles. 
To adjust for time-dependent covariates over a relatively long period of time, requires one to observe a high number of covariates that are frequently unknown when the study is designed. 
For example, external factors such as cost of fuel, traffic, and parking fare attributes to car owners' preferences on the travel mode \cite{Gao2021}, and most of these factors are challenging to observe from the perspective of the vehicle.
Thus, in a longitudinal software evaluation, there is a need for models that can control covariates even when they are not observed.

Study II is designed to explore the scenario described above, that is, (a) when the performance of the target software treatment is highly dependent on external and seasonal factors such as temperature, (b) and there are potentially latent variables that cannot be observed in an effective manner such as travel demands of individuals. 
The study is designed to assess the applicability of the BDID model in addressing the challenge and to verify the causal assumption in BDID has real-life relevance. The software deployed in study II is a battery management system for electric vehicles. 
The battery inside of an electric vehicle has an ideal window of operating temperatures, at the start of a trip, the battery management system will have to either warm up or cool down the battery into said window of operation, this is done at the cost of driving range. 
Therefore, the ideal preconditioning operation shall take place during charging prior to the trip, utilising the electricity from the grid instead of from the battery. 
Moreover, if there is a large difference between the ambient temperature and ideal operating temperature, the energy required to precondition the battery is naturally higher. In other words, it is reasonable to expect a dynamic seasonality effect on the final average treatment effect. The treatment is a software solution that controls and optimises the precondition of battery, and it is expected to improve range as the vehicle no longer needs to heat up or cool down the battery during driving operations. Therefore, it decreases the energy consumption (measured in $\mathrm{Wh/km}$), and if the software performs as expected, the average treatment effect should lower energy consumption and extend range.

The study took place between the 1st of August 2021 to the 30th of September 2021, note that during our study, there is a two-week duration that is a typical vacation period in Sweden where the vehicles' users reside. We choose to conduct the study during this period as the weather conditions are dynamic as well as the travel demands, to further demonstrate the power of the BDID model. During the period, we collected data from 24,286 trips and a total of 616,212 kilometres. The control group, similarly to study I, is running the existing version of the software and we do not intervene with the vehicles besides passive data collection. 
The treatment group is randomly sampled from a larger fleet of vehicles leased to company employees, and these vehicles received the battery preconditioning software as described above. There are in total 176 vehicles included in the study, similarly to study I, we discard data generated by vehicles with odometer less than 100 kilometres and all trips with average speed higher than 200 kilometres per hour as it exceeds the digital speed limiter implemented in the vehicles. We have in total 9 covariates, and their descriptive statistics are presented in Table. \ref{table_data_bdid}. All values are presented min-max scaled due to nondisclosure agreement with our industry collaborator.

\begin{figure}[t]
\centerline{\includegraphics[width=\linewidth]{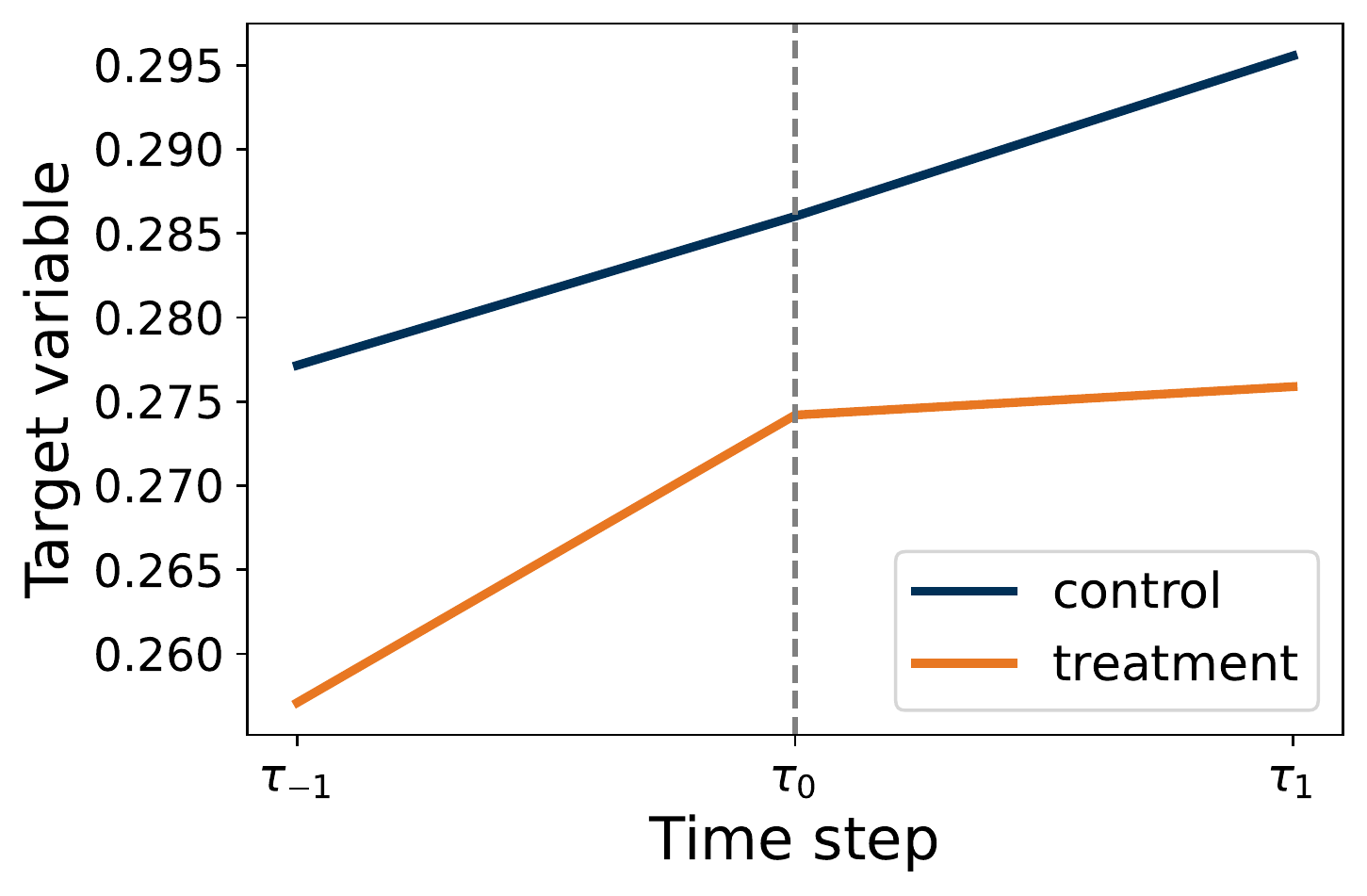}}
\caption{Target variable $y$ measured before the treatment ($\tau_{-1}$), when the treatment is applied ($\tau$), and after the treatment ($\tau_1$), for the control and the treatment groups, $y$ is min-max scaled.}
\label{fig_DID_parallel}
\end{figure}

While the Bayesian difference-in-differences model essentially allows a \textit{post facto} analysis of time dependent treatment effect analysis, it is based on the assumption of parallel trend, as formally defined in Equation \ref{eq:parallel}. 
In practise, this trend states that the control and the treatment group should follow similar trend over time, if no treatment is applied, implying the difference in between the groups comes from the unobserved time dependent confounding factors. 
For BDID to identity the treatment effect, the assumption needs to be empirically validated through for example visualisation, i.e., by comparing the target variable over time between the treatment and the control group before an intervention is introduced. Another validation method is to fit the BDID model before and after the treatment is applied, to test if the functional form of the counterfactual is correct \cite{KahnLang2019}.
Empirically, some observe the samples pre-treatment for as long as possible, to discovery any unknown or underlying trend over time \cite{Angrist2010}.
Furthermore, some matching is required when selecting the control group to compare with the treated group. In practise, this matching process can be done by selecting untreated samples that are as similar as possible to the treated samples, such as vehicle model and engine types, markets, and so on. To ensure the two groups are comparable, we select vehicles with the same vehicle type and have the same battery capacity, and all of the vehicles are registered and driven in Sweden. 

\subsection{Results}

In this subsection, the results from study II is presented. We show the BDID regression model inferred from a Bayesian network, we illustrate the parallel trend assumption, and the final average treatment effect of the software change. 

First we inspect the parallel trend assumption through visualisation and the result is presented in Fig. \ref{fig_DID_parallel}. First, we compute the average of the target variable before a treatment is applied, at discrete time step $\tau_{-1}$ for both control and treatment groups, and the target variable at the time when the new software is introduced, $\tau$. As can be seen from the figure, the target variable from both groups follow a upward trajectory. Last, we compute the average of the target variable after the treatment is applied at discrete time step $\tau_1$. The visualisation result confirms our assumption that both the control and treatment groups follow similar trend over time, if no treatment is applied, and the target variable observed in the treatment group changes trajectory after the treatment application.

\begin{figure}[t]
\centerline{\includegraphics[width=\linewidth]{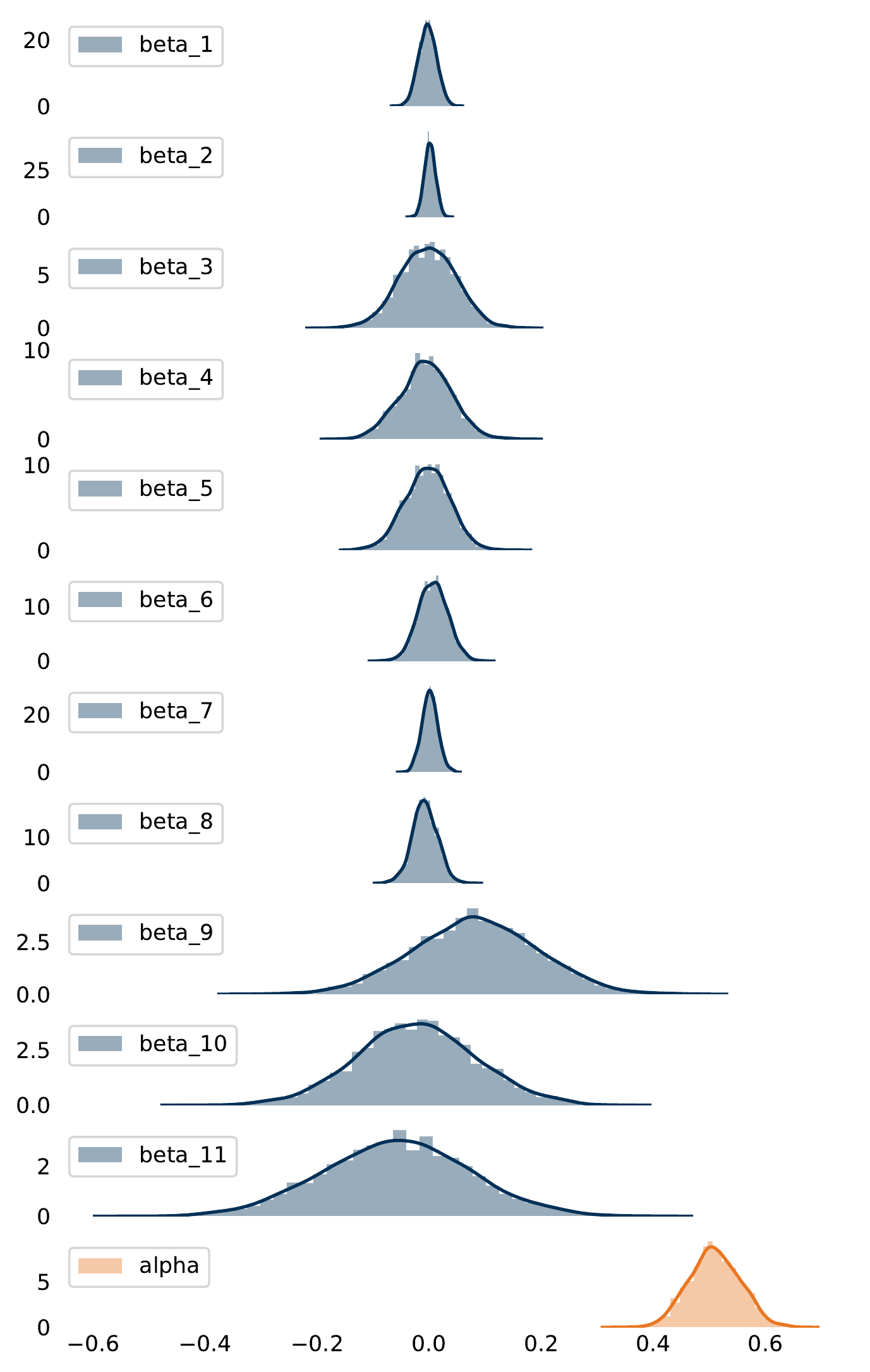}}
\caption{Posterior distribution of the regression coefficients $\beta$, $\beta$ is ordered as Table. \ref{table_data_bdid}, and the regression intercept $\alpha$.}
\label{fig_BDID_post}
\end{figure}

The BDID regression is implemented in Pyro, similarity to the Bayesian logistic regression in study I. We follow the generative process as prescribed in Algorithm \ref{bdid_gen}. The MCMC NUTS sampler is set in Pyro with 3.000 samples and 200 burn-ins with two chains. The Brooks-Gelman-Rubin convergence criteria of $\hat{R} < 1.1$ is met ($\hat{R} = 1.0007$ for $\alpha$, $\hat{R}_{\beta, mean} = 1.0066$ for all $\beta$, and $\hat{R} = 0.999$ for the error term $\sigma$). We attach the trace plots in the online appendix.

\begin{table}[t]
\caption{Average energy consumption ($\mathrm{Wh/km}$) for the control and the treatment group at each time step.}
\centering
\begin{tabular}{p{0.07\textwidth}p{0.1\textwidth}p{0.1\textwidth}p{0.1\textwidth}}
\hline
 & Control & Treatment & Difference  \\ \hline\hline
$\mathrm{\tau_{-1}}$ & 205.30 & 190.45 & - 14.85\\ \hline
$\mathrm{\tau_{1}}$ & 218.93 & 204.36 & - 14.57\\ \hline
Change & 13.627 & 13.915 & \textbf{-0.280}  \\ \hline
\end{tabular}
\label{tabel_bdid}
\end{table}

We assign a weakly informative prior distribution of $\alpha \sim\mathcal{N}(0, 1)$ and $\beta \sim\mathcal{N}(0, 1)$ to the regression intercept and coefficients to introduce scale information to regularise inference, in this case, min-max scalded covariates. We infer the posterior distribution $p(y, \alpha, \beta | \mathbf{X}, \sigma, \lambda_{\alpha}, \lambda_{\beta})$, in Fig. \ref{fig_BDID_post}, we present the posterior distribution of the regression intercept ($\alpha$) and the regression coefficients $\beta$. 
As can be seen, the posterior distribution of the two dummies variables indicating the time period and if a treatment is applied ($\beta_1$ and $\beta_2$) are informative of the treatment outcome as expected. 
While the posterior distributions for covariates describing the high voltage battery activity level such as total energy discharge and state-of-charge change ($\beta_9$ to $\beta_{10}$), contribute positively to average energy consumption, however, the uncertainty of the effect is high. 

Last, we include a difference in differences average treatment effect analysis following Equation. \ref{eq:ATE_DID}. First, we compute the difference of expected target variable $\mathbb{E}[y]$ at time step $\tau_{-1}$ between the control ($t=0$) and the treatment ($t=1$) groups, this value can be interpreted as the preexisting differences between the groups as a result of unobserved confounding effects. Second, we calculate the difference of $\mathbb{E}[y]$ between the control and treatment group at time step $\tau_1$, this difference is a sum of the preexisting differences and the treatment effect if the parallel assumption holds. The results are presented in Table. \ref{tabel_bdid}.

\section{Bayesian regression discontinuity design \label{BRDD}}

In this section, we present the Bayesian regression discontinuity design (BRDD) theory and observational study III that is designed to demonstrate the use case of BRDD for evaluating automotive software. The theory and assumptions of the model is presented formally along with the algorithm for Bayesian inference. We present the study III, the setup, data collection method, and how BRDD is used as a strategy for identifying continuous covariate dependent treatment assignment.

\begin{figure}[t]
\centerline{\includegraphics[width=\linewidth]{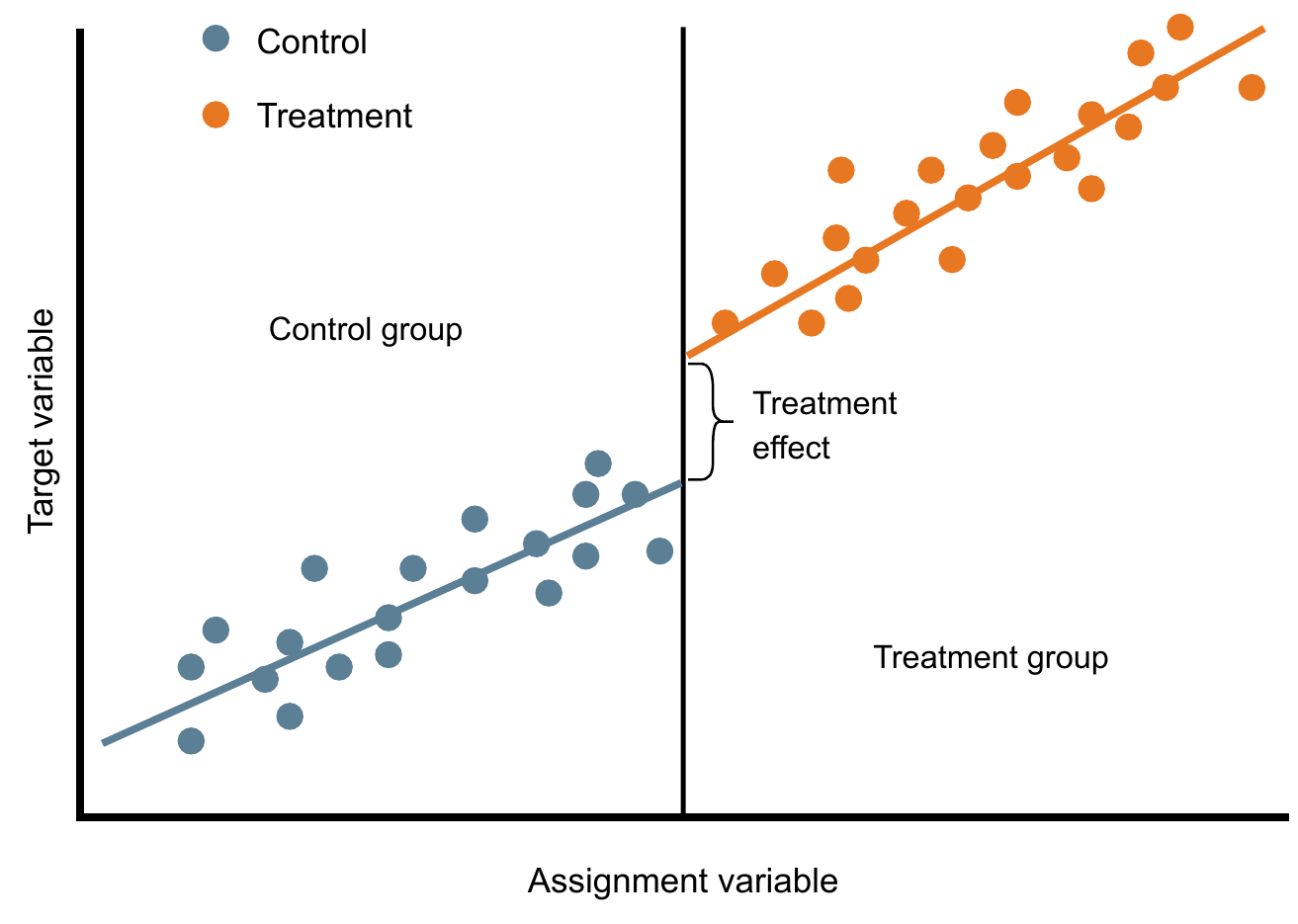}}
\caption{An illustration explaining the Regression Discontinuity Design model, and how the average treatment effect is estimated.}
\label{fig_RDD}
\end{figure}

Regression discontinuity design, proposed by \cite{Thistlethwaite1960}, is a causal modelling approach aiming to determine the treatment effect when the treatment assignment is confounded by one continuous covariate by assigning a cut-off point. This continuous covariate is referred as the assignment variable. In Fig. \ref{fig_RDD}, we illustrate the principle of RDD. Observations of the target variable is made around the threshold, in this case, average treatment effect can be inferred without the need of a randomised experiment. In practise, visualising of the assignment variable and the target variable is a simple yet powerful tool to inspect their relationship \cite{Imbens2008}. 
While RDD allows inference of treatment effect with the absent of randomisation, the model alone does not explicitly or implicitly conclude causality as it does not identify other unobserved confounding effects. Moreover, the RDD model essentially infer the treatment effect with a single covariate $X = x$, in that perspective, the model has a limited degree of external validity. However, the RDD model is similar to a randomised experiment with bias below 0.01 standard deviations on average especially analysed with the Bayesian approach \cite{Chaplin2018}, indicating a high internal validity. 
In our study III, we investigate the scenario when a software is only useful in reducing the fuel consumption of the vehicle, if the average trip distance of the given vehicle is over a certain threshold, without \textit{ex-ante} randomisation. 

\subsection{Theory}

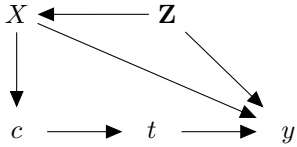
\begin{figure}[t]
\centering
\begin{tikzpicture}[
roundnode/.style={circle, draw=black!60, fill=white, very thick, minimum size=10mm},
roundedrect/.style={rectangle, rounded corners, minimum width=3cm, minimum height=1.5cm,text centered, draw=black}
]
\node      (maintopic)          {\begin{tabular}{c}$t$\end{tabular}};
\node      (rightcircle)       [right=of maintopic] {\begin{tabular}{c} $y$\end{tabular}};
\node      (cutoff)            [left=of maintopic] {\begin{tabular}{c} $c$\end{tabular}};
\node      (uppercircle)       [above=of cutoff, xshift=2cm] {$\mathbf{Z}$};
\node      (assignment)        [above=of cutoff] {$X$};

\draw[->] (uppercircle) -- (rightcircle);
\draw[->] (uppercircle.west) -- (assignment.east);
\draw[->] (maintopic.east) -- (rightcircle.west);
\draw[->] (cutoff.east) -- (maintopic.west);
\draw[->] (assignment.south) -- (cutoff.north);
\draw[->] (assignment) -- (rightcircle);

\end{tikzpicture}
\caption{A simplified directed acyclic graph showing the relationships of treatment ($t$), target variable ($y$), assignment variable ($X$), the cut-off point ($c$), and other confounding factors ($\mathbf{Z}$).}

\label{figure:DAG_RDD}
\vspace{-0.5cm}
\end{figure}

In \cite{Thistlethwaite1960}, RDD is discussed in the context of regression, and in this subsection, we will describe it in the potential outcomes framework provided the conditional exchangeability assumption holds as formulated in Equation. \ref{eq:conditional_exe}.
We illustrate the relationship between variables in a RDD in Fig. \ref{figure:DAG_RDD}. As can be seen, the outcome $y$ is influenced by the assignment variable as well as the predetermined cutoff point $c$. As mentioned previously, a RDD does not automatically eliminate other confounding factors in the system, we illustrate that with $\mathbf{Z}$ in the directed acyclic graph.
The fundamental concept of RDD is that the treatment assignment is deterministic by a covariate $X$ (we call this the assignment variable) with a fixed threshold, the assignment variable $X$ is assumed to be correlated to the target variable $y$, and their correlation is smooth. Under this assumption, any discontinuity of the target variable $y$ as a function of $X$ is interpreted to be the causal treatment effect around the predetermined threshold. Let $X=c$ be the predetermined cut-off point of the assignment variable with $c$ being an arbitrarily determined value, formally,

\begin{equation}
     \mathbb{E}[y|X=x, t=1], \: \mathrm{and} \; \mathbb{E}[y|X=x, t=0]
     \label{ass_did}
\end{equation}

\textit{are continuous in} $x$.

This assumption also implies that the probability distribution of $y$ given the covariate $X=x$ is smooth.
This assumption is stronger than needed, as the continuity without a treatment effect is only expected around the cut-off point $X=c$ and the assumption above covers such a scenario.
Different from a matching problem, the requirement for overlap requires control and treated samples to have all possible combinations of the covariates, in a DID model, for all values of $x$, the propensity of treatment assignment is either $0$ or $1$, i.e., on either side of the cut-off point $c$. We call this a sharp design, as opposed to fuzzy design. In practise, a fuzzy design might be more attractive, as it is reasonable to include samples close to either side of the cut-off point.

Without the need of extrapolating due to the lack of overlap, at the cut-off point $X=c$, we can infer the average treatment effect from regression discontinuity design ($ATE_{RDD}$) as,

\begin{equation}
    ATE_{RDD} = \mathbb{E}[y|X=c, t=1] - \mathbb{E}[y|X=c, t=0]
    \label{eq:ATE_rdd}
\end{equation}

That is, the difference between average observed target variable $y$ with or without the treatment $t=\{0, 1\}$, at a given cut-off point $X=c$. This average treatment effect can be estimated as we have made the smoothness assumption in Equation. \ref{ass_did}. 
We would like to empathise that although we demonstrate a linear regression for the prediction of target variable in a BRDD, a polynomial regression can be applied to handle more complex relations between the assignment variable and the target variable. 
When the smoothness assumption holds, the target variable $y$ can be predicted using a simple linear regression for the sharp design at $X=c$, stated as following centred around the cut-off point,

\begin{equation}
    y \sim \alpha + \beta_{1}(x - c) + \beta_{2}t + \beta_{3}(x - c)t + \beta_{4}Z + \epsilon
\end{equation}

where, the $\alpha$ is the regression intercept, $\beta_{j}$ are the regression coefficients. They are both latent variable inferred from a Bayesian network. $t$ is a dummy variable indicating if a treatment has been applied, and $\epsilon$ represent the linear noise in the model. 
We assign a Gaussian distribution as a prior to the regression intercept and coefficients, namely, 

\begin{equation}
    \alpha \sim \mathcal{N}(\alpha|0, \lambda_{\alpha}) 
\end{equation}

and,

\begin{equation}
    \beta_{j} \sim \mathcal{N}(\beta_{j}|0, \lambda_{\beta_{j}}) 
\end{equation}
 
An error term $\epsilon$ is included in the model which represent the observation noise as a linear model has its limitations for describing the noisy reality. We have,

\begin{equation}
    \epsilon \sim \mathcal{N}(\epsilon | 0, \sigma^2)
\end{equation}

The joint distribution is factorised as the blow applying Baye's law. We describe the generative process for this join distribution in Algorithm \ref{brdd_gen}.

\begin{equation}
\begin{aligned}
    &p(y, \alpha, \beta_j | Z, t, x, c, \sigma, \lambda_{\alpha}, \lambda_{\beta_j}) \\
    &= p(t) \cdot p(\alpha|\lambda_{\alpha}) \cdot  p(\beta_j|\lambda_{\beta_j}) \cdot \prod_{n=1}^{N}p(y|Z, t, x, c, \alpha, \beta_j, \sigma)
\end{aligned}
\end{equation}

\begin{algorithm}[t]
\caption{Bayesian regression discontinuity design generative process}
\textbf{Inputs:} $\mathbf{X}$ covariates, $\lambda_{\alpha}$ prior distribution of $\alpha$, $\lambda_{\beta}$ prior distribution of $\beta$, treatment effect $y_n$ 
\begin{algorithmic}[1]
\State Draw $\alpha \sim \mathcal{N}(\alpha|0, \lambda_{\alpha})$
\For{each $\beta \in \beta_j$}
    \State Draw $\beta \sim \mathcal{N}(\beta|0, \lambda_{\beta})$
\EndFor
\State Draw $\epsilon \sim \mathcal{N}(\epsilon|0, \sigma^2)$
\State Draw $y \sim \mathcal{N}(y|\alpha + \beta_1(x-c) + \beta_2t + \beta_3(x-c)t + \beta_{4}Z, \sigma^2)$

\end{algorithmic}
\label{brdd_gen}
\end{algorithm}
 
\subsection{Study III: Covariate dependent treatment assignment}

In absence of randomisation, assuming automotive software functions to be independent from their operation environment or usage by the customers is a naïve approach for estimating the treatment effect of software changes. To that end, the performance or even the activation of a certain automotive function is dictated by the usage, in other words, we frequently run into the situation in which a covariate determines the treatment assignment. 
To understand the performance of this type of software is important for the following two reasons. First, it brings insights on the usefulness of a given software feature, validating assumptions made during development against how the product is actually utilised. Second, it allows the development organisations to evaluate the software effectiveness in conditions that are most determining of the effect. In study III, we present a case that illustrates the importance of causal inference when the treatment effect is strongly dependent on a covariate.

A plug-in hybrid vehicle, is a type of electrified vehicle with two sets of propulsion, combustion engine and electric motors. 
This type of vehicles usually have limited pure electrical range, and to maximise the benefit of eclectic drive such as high efficiency for low speed driving and zero direct emission; automotive manufacturers typically have a number of control software solutions to optimise the distribution of the electrical and chemical energy on a given trip. 
A simple version of the optimisation strategy is to prioritise the electrical energy whenever available and deplete the battery first before using the combustion engine. This type of strategy usually works well when the driver is expected to travel less distance than the electrical range, and not on highways where the combustion engine works more efficiently than the electrical motor. 
Alternatively, the optimisation can be done through a prediction of trip distance and destination -- if the trip is predicted to be farther than the electrical range, the car will not prioritise the use of electrical energy and deplete the battery early in the trip, with the rationality that the drivers is predicted to enter the city later where direct emission from the combustion engine is undesirable. 
Thus, the assignment of this software is determined by the trip distance by design.
The performance of this type of software function is highly dependent on how the cars are driven, more specifically, on the trip distance. 
Thus, to evaluate such a software feature, we chose a cutoff point of the assignment variable trip distance, at around the designed electrical range of the vehicle where the software is expected to have the most impact, and apply the BRDD model for treatment effect inference and modelling. 

Study III is conducted in Sweden, on a fleet of 50 plug-in hybrid vehicles. The study took place between the 19th of October 2020 to the 28th of Febuary 2021, and during which, 12,231 numbers of trips are observed and the vehicles have driven a total of 191,552 kilometres. 
The software is designed so that if the predicted trip distance is less than the electrical range, the car will prioritise and use the electrical energy first. If not, the vehicle will optimise the energy usage between the electrical motor and the combustion engine according to the predicted trip distance and destination. 
We apply the same data post-processing logic as study I and II, namely, data collected from brand new vehicles and trips with higher than possible average speed are excluded from the model.
In study III, the target variable is the average fuel consumption, and the assignment variable is the total trip distance.

There are two important assumptions in a regression discontinuity design. First, we assume unobserved covariates do not affect the treatment effect or assignment. This is a strong assumption, similarly to what is previously discussed for BPSM and most causal inference problems, to satisfy this assumption, it requires prior knowledge to how the software interact with the users and inputs from the domain experts. The second assumption of BRDD is the expectation of the target variable $\mathbb{E}[y]$ is continuous with respect to the assignment variable $X$. 
Mathematically speaking, function $f(x)$ continuity at $x=c$ can be determined if $\lim_{x \to c} f(x)$ exist, and $\lim_{x \to c} f(x) = f(c)$. 
A continuity check should not be performed on observational data which is per definition discontinuous, instead, the continuity assumption can be check with a density test, as suggested by \cite{McCrary2008}.
Last but not least, the choice of cut-off point $X=c$, requires that the assignment mechanism to be known to the development teams. In practise, the cut-off point might not be a sharp differentiation but rather a bandwidth, which can be determined either through the design intend of the software or through observational data collected prior to a treatment is introduced.

\subsection{Results}

\begin{figure}[t]
\centerline{\includegraphics[width=\linewidth]{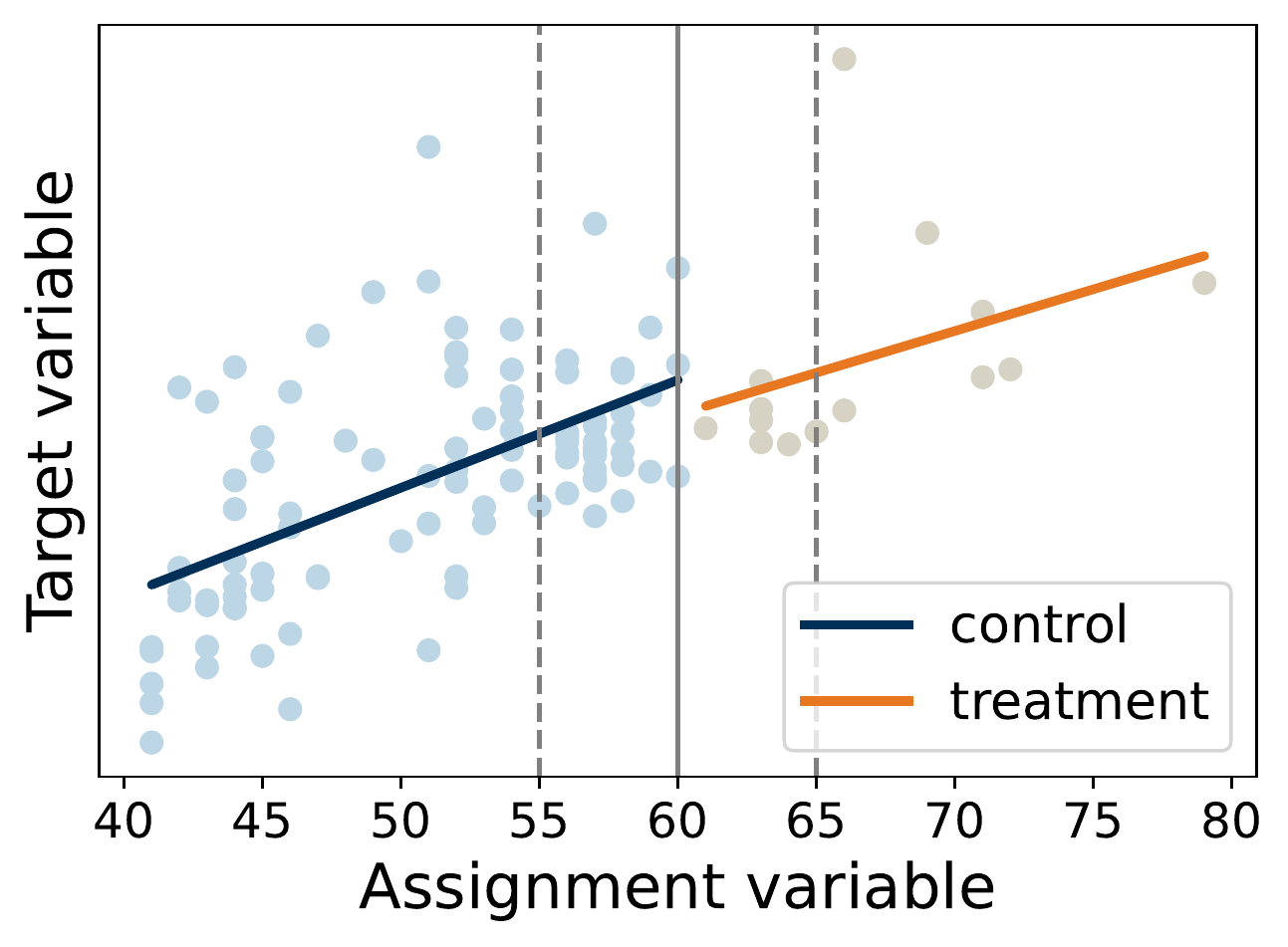}}
\caption{Target variable measured before and after the cut-off point ($c=60$), with respect to the assignment variable.}
\label{fig_rdd_wdata}
\end{figure}

In this section, we present the results from BRDD. As discussed in the previous subsection, there are other covariates that could potentially confound the treatment effect, in $\mathbf{Z}$. 
In this analysis, to adjust for the covariates, we condition on one covariant that is the total displaced state-of-charge, $Z \in \mathbf{Z}$. This covariate indicates how much battery is used during a given trip, naturally, if a trip distance is fixed, the more battery is used, the less the fuel consumption there is. 
Thus, to ensure the trips are comparable disregard the software treatment, we only look at $\mathbb{E}[y|Z > 90]$, they are trips during which the battery has been depleted.
The cut-off point is arbitrarily determined at $c=60$, which is the approximated pure electrical range of the vehicle model we are observing. We do not min-max scale the value in Fig. \ref{fig_rdd_wdata}, due to that the cut-off point of the assignment variable represents a physical measurement and we choose to articulate the physical meaning through presenting the unscaled value. As a min-max scaled value is difficult to interpret, to reflect the physical meaning of the measurement while maintaining our confidentiality agreement, we remove the units of the measured target variable instead. 
Two linear regression is fitted on either side of the cut-off point, to illustrate a discontinuity of the regression line at cut-off as a representation of the software treatment effect, as can be seen in Fig. \ref{fig_rdd_wdata}, at $c=60$.

\begin{figure}[t]
\centerline{\includegraphics[width=\linewidth]{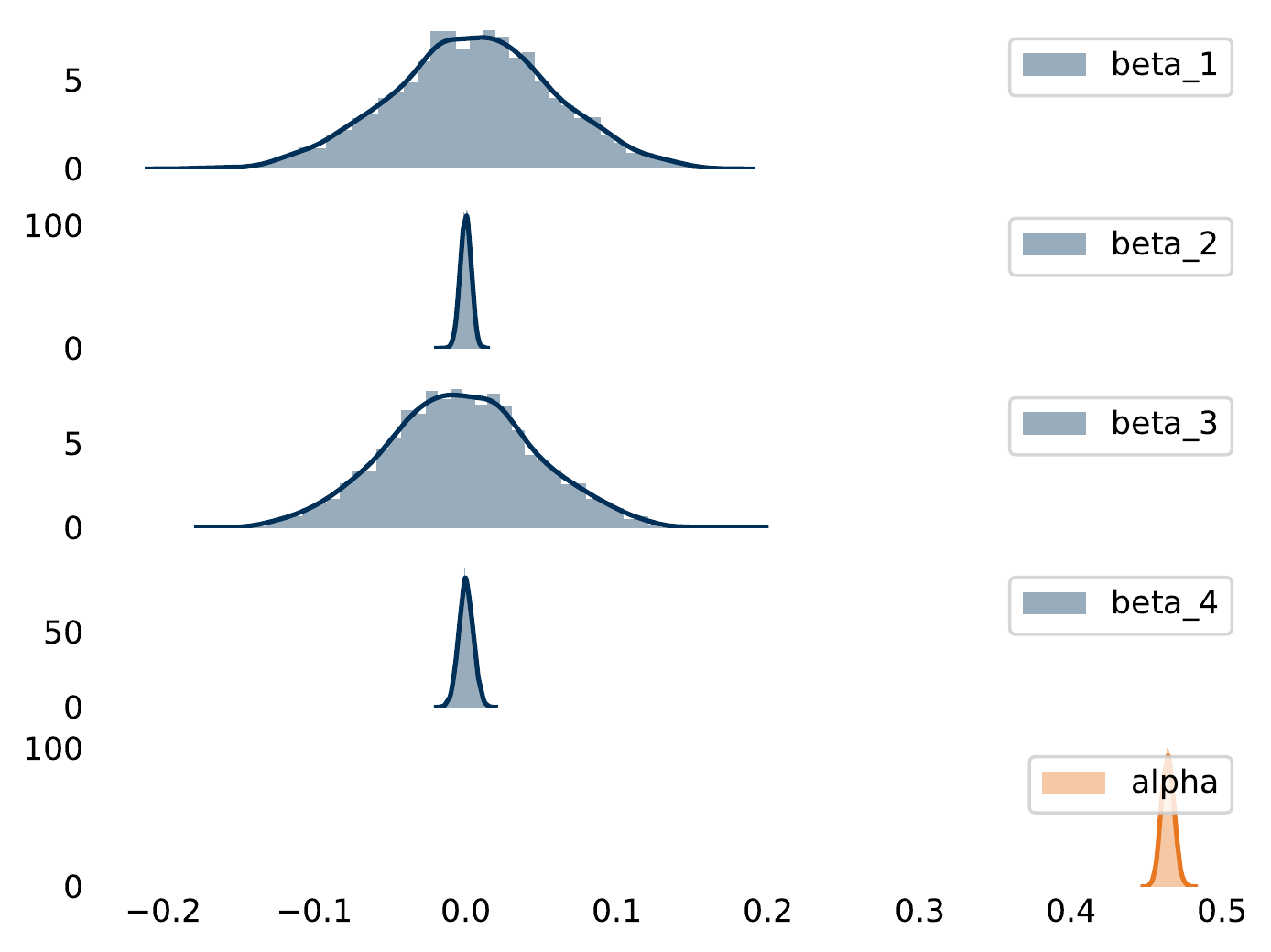}}
\caption{Posterior distribution of the regression coefficients $\beta$, and the regression intercept $\alpha$.}
\label{fig_BRDD_post}
\end{figure}

Following Algorithm \ref{brdd_gen}, we implement the BRDD regression model in Pyro. The MCMC NUTS solver is set with 2 chains of 2,000 samples each and we discard the first 200 steps as warm-up steps. 
The model, in wide format, has the following four input features, $x-c$ (so that the regression is fitted centred around the cut-off point), with $x$ being the assignment variable of trip distance and $c=60$, $t = \{0, 1\}$ (to effectively control the regression model), $(x-c)t$ ($(x-c)t=0$ for the controlled group, and $1$ otherwise), and the change of the state-of-charge of the battery named $Z$. 
The convergence criteria $\hat{R} < 1.1$ is met at $\hat{R} = 1.0004$ for $\alpha$, $\hat{R}_{\beta, mean} = 1.0017$, $\hat{R} = 1.0000$ for the error term $\sigma$. The trace plots are attached in the online appendix. 
For the prior distributions, we have a weakly informative prior of $\alpha \sim \mathcal{N}(0, 1)$, and we select a more informative prior for $\beta \sim\mathcal{N}(0, 0.5)$. We choose a weakly regularising prior for the standard deviation $\sigma \sim \text{Half-Cauchy}(0, 5)$, as it approximate uniform distribution and it is weakly informative near 0. 
We plan to start with a non-informative prior for the error term and adjust if the solver does not converge. The solver meets the convergence criteria with the priors mentioned above.
The posterior distribution from this Bayesian network, $p(y, \alpha, \beta |Z, t, x, c, \lambda_{\alpha}, \lambda_{\beta}, \sigma)$, is inferred. The posterior distribution of the regression intercept, $\alpha$, returned a Gaussian distribution centred around $\alpha_{mean} = 0.626$, with a standard deviation of $\alpha_{std}=0.0056$. Similarly, the posterior distribution of the error term $\epsilon$ is a Gaussian distribution centred around $\sigma = 0.213$ with standard deviation of $0.0010$. We present the posterior distributions in Fig. \ref{fig_BRDD_post}.

The outcome for BRDD consists of two linear regression lines at either side of the cut-off point. Using the posterior distribution, the regression expression ($y_{c}$) to the left hand side of the cut-off point can be expressed as,

\begin{equation}
    y_{c} \sim \alpha + \beta_1x + \beta_4Z + \epsilon
\end{equation}

and the regression expression for the treated samples ($y_{t}$) to the right hand side of the cut-off point can be expressed as,

\begin{equation}
    y_{t} \sim (\alpha + \beta_2) + (\beta_1 + \beta_3)x +\beta_4Z + \epsilon
\end{equation}

The average treatment effect $ATE_{RDD}$ is inferred following Equation. \ref{eq:ATE_rdd}, we find that, at the cut-off point ($X=c$), the expected target variable treated and untreated differ by $ATE_{RDD} = y_t(x=c) - y_c(x=c) = -1.1954$. This is an unbias estimation of the local conditional treatment effect. As can be seen in Fig. \ref{fig_rdd_wdata}, the discontinuity at the cut-off point can be interpreted as the treatment effect. 
\section{Discussion \label{diss}}

In this section, we provide a discussion on the advantages and limitations of the BOAT framework from the perspective of causal inference and piratical applications in the automotive domain. 

\subsection{Causal assumptions and domain knowledge}

Causality cannot be inferred from observational data alone \cite{Pearl2009}, as behind every causal conclusion, there are some complementary causal assumptions that might not be testable empirically. 
Before adjusting for confounding biases, some judgements must be made based on domain knowledge as also discussed by \cite{Rubin2001, Pearl2009, Xu2009, Stuart2010}.

Take propensity score matching as an example, different from a randomisation process, in which all covariates will be balanced in the control and the treatment groups observed or not, causal inference with covariate adjustment requires the covariates to be observed \cite{Stuart2010}. Performing propensity score matching in combination with observational study, requires a set of carefully chosen covariates which rely heavily on domain knowledge.
To that end, similar requirements on domain knowledge is also experienced with other models in the BOAT framework. 
In order to select samples in the control group for observation that are as similar as the treated samples as possible, a matching process is recommended \cite{KahnLang2019}. The judgement on "as similar as possible" is a judgement that cannot be made without existing data on the cohorts and domain knowledge. 
Likewise, domain knowledge is required for selecting the cut-off point for the assignment variable in a regression discontinuity design.

The requirement on domain knowledge implies that causal inference from observational studies need manual input when applied in software engineering practises. 
While randomised online experiments can be automated to a large extend, as demonstrated in the SaaS domain \cite{google2010, Deng2013, Xie2016}, causal inference with observational data is a process that would potentially require more manual efforts. 
The extra overhead of efforts from developers could potentially pose a challenge in implementing causal inference in combination with observational studies in automotive software engineering. 
Thus, the application of BOAT framework on a larger scale cannot be done without some form of data-driven causal discovery methods.



\subsection{Extension to BOAT}

In this subsection, we offer a short discussion on the potential outlook of the BOAT framework.
As shortly discussed in Section. \ref{theory}, the BOAT framework does not cover all scenarios in causal inference of observational studies in the automotive domain.
Since there are a few more techniques that can be applied when inferring causal treatment effect without randomised experiments.
In this subsection, we offer a discussion on our decision to why some of the models are not included in the BOAT framework as if now.
These models, useful in many domains as literature reports, we have yet to find their feasible applications in the automotive sector.

First, the positive decision to whether there is a need to infer the latent variable is intentionally left out from the flow chart, when a latent variable needs to be inferred, methods such as instrumental variable can be applied \cite{Imbens1994}. Instrumental variable method uses a latent variable to explain the correlation to the error term. 
The method should account for unexpected behaviour between variables, however, the explanation it provides cannot be interpreted with a physical meaning. In many automotive software where interpretability is considered crucial, especially for development organisations to take design decisions. Moreover, instrumental variable method has the tendency to produce bias results when the sample is small, which is a known limitation in the automotive domain. 

Second, a popular school of causal inference method, structural causal model and do-calculus \cite{Pearl2009} offers a comprehensive approach to causal inference provided the causal structure is known. This causal structure is represented in a DAG, such as the trivial example in Fig. \ref{figure:DAG}. 
Each component in a DAG has their graphical and numerical representations, then through the language of do-calculus, for example $p(Y|do(T), X)$, we can represent intervention and infer treatment effect from a DAG. 
In order for a structural causal model to be effective, the structure of the model, i.e., the DAG, needs to be learnt either through domain-knowledge or though a data-driven causal discovery process. The formal is time consuming and potentially subjected to biases of individuals, the latter requires large amount of data yet does not address limitations such as sampling bias, measurement error, and confounding effects \cite{Glymour2019}.

Finally, there are other methods addressing preexisting differences between the control and the treatment groups, such as inverse propensity weighting. While achieving similar objective as propensity score matching (adjusting for confounding factors), inverse propensity weighting is a parametric method and it is known to be creating imbalance groups when the sample size is insufficient.

\section{Conclusion \label{conclude}}

In this paper, we introduce the BOAT framework for software engineering in the automotive domain, enabling online evaluation of the software in a causal fashion when a fully randomised experimentation is impossible, undesired, or unethical. 
Applying the Bayesian causal inference models, we demonstrate how a causal conclusion can be drawn in absent of randomisation utilising the high flexibility of Bayesian inference towards sample size, as demonstrated in other areas of science \cite{Normington2019, Li2020, Lee2008, Chib2015}.
Combining theory with practise, we include three illustrative cases from the automotive domain for further enforce the need of causal inference in software engineering. The three cases are conducted together with our industry collaborator, we introduce three software to a fleet of vehicles driven by real-world customers.
We relate the causal assumptions to scenarios experienced in practise, aiming to provide a guideline on when and how to better apply the causal modelling for inferring the software effects from different software evaluation needs.

We provide a decision making flowchart along with the three causal inference models included in the BOAT framework. The flowchart is design with the objective of guiding development organisations on which causal inference models should be used to address their corresponding challenges in real life. 
In the BOAT framework, we include three models, they are, (1) Bayesian propensity score matching for generating balanced control and treatment groups without randomisation, (2) Bayesian difference-in-differences for controlling unobserved seasonal factors over time, (3) Bayesian regression discontinuity design for analysing the treatment effect when treatment assignment is determined by a continuous covariate. 
All of the three models and their assumptions have their implications in practise, as experienced from the automotive domain when attempting to evaluate software without randomisation, in this work, we provide a formal discussion of the inference models, as well as their corresponding real world implications and applications. 
The three cases are designed together with software engineering teams in our case company, to simulate challenges experienced when evaluating software online without randomisation.
With the development teams, we introduce new software treatment to a fleet of vehicles, conduct data collection, and use the empirical data as inputs to the BOAT framework, additionally, we assess the causal assumption in relation to the empirical cases.
We find the causal models in the BOAT framework to be highly applicable in automotive software engineering, and they enable the development organisations to evaluate software changes in an online and causal manner. 
Furthermore, we find there is a strong dependency on domain knowledge when designing an online observational study as the cause-and-effect is not always known, and when validating some of the causal assumptions empirically.

In our future work, we aim to incorporate causal discovery process when designing an online experiment or observational study, since we cannot always assume the cause-and-effect of a system is known \cite{Scholkopf2021}. Moreover, we plan to explore data-driven causal discovery methods to inform and potentially automate the design of experiments, with the objective of increasing the efficiency and the effectiveness of online experimentation in automotive software engineering. 

\section*{Online Appendix \label{app}}

We attached an online appendix for the Bayesian models.
The online appendix can be found as a Jupyter Notebook via the following link: \href{https://github.com/yuchueliu/BOAT}{github.com/yuchueliu/BOAT}.

\section*{Acknowledgement}
This work is supported by Volvo Cars, by the Swedish Strategic vehicle research and innovation programme (FFI), and by Chalmers University of Technology.

\bibliographystyle{IEEEtran}
\bibliography{IEEEabrv, ref.bib}

\end{document}